\documentclass[review,3p]{elsarticle}

\usepackage{lineno,hyperref}
\usepackage{graphicx}
\usepackage{graphics}
\usepackage{subfigure}
\usepackage{amsmath}
\usepackage{multirow}
\usepackage{amsfonts}
\usepackage{soul}
\usepackage{color, xcolor}
\soulregister{\cite}7  
\soulregister{\ref}7   
\soulregister{\eqref}7  
\soulregister{\pageref}7   
 
\usepackage{booktabs}

\begin{document}

\begin{frontmatter}

\title{SPNet: A novel deep neural network for retinal vessel segmentation based on shared decoder and pyramid-like loss}

\author[mainaddress]{Geng-Xin~Xu}
\author[mainaddress,secondaryaddress]{Chuan-Xian~Ren\corref{correspondingauthor}}
\cortext[correspondingauthor]{Corresponding author}
\ead{rchuanx@mail.sysu.edu.cn}

\address[mainaddress]{School of Mathematics, Sun Yat-Sen University, Guangzhou, 510275, China}
\address[secondaryaddress]{Key Laboratory of Machine Intelligence and Advanced Computing (Sun Yat-Sen University), Ministry of Education, China}

\begin{abstract}
Segmentation of retinal vessel images is critical to the diagnosis of retinopathy. {Recently, convolutional neural networks have shown significant ability to extract the blood vessel structure. However,} it remains challenging to refined segmentation for the capillaries and the edges of retinal vessels due to thickness inconsistencies and blurry boundaries. In this paper, we propose {a novel deep neural network for retinal vessel segmentation based on shared decoder and pyramid-like loss (SPNet)} to address the above problems. Specifically, we introduce a decoder-sharing mechanism to capture multi-scale semantic information, where feature maps at diverse scales are decoded through a sequence of weight-sharing decoder modules. Also, to strengthen characterization on the capillaries and the edges of blood vessels, we define a residual pyramid architecture which decomposes the spatial information in the decoding phase. A pyramid-like loss function is designed to compensate possible segmentation errors progressively. Experimental results on public benchmarks show that the proposed method outperforms {the backbone network} and the state-of-the-art methods, {especially in the regions of the capillaries and the vessel contours}. In addition, performances on cross-datasets verify that SPNet shows stronger generalization ability.
\end{abstract}


\end{frontmatter}

\section{Introduction}
\label{sec:Introduction}
Retinal vessel is an important part of microcirculation in human body. Its morphological changes in diameter, tortuosity and other characteristics are closely related to the severity of ophthalmological diseases and cardiovascular events, e.g., glaucoma, diabetes, and hypertension \cite{cheung2017imaging, yu2020framework, wu2020nfn+, ma2019multi}. In order to analyze the changes in vascular morphology, specialists usually segment the vessels from fundus images manually. However, manually extracting vessels is time-consuming and may lead to different results among experts due to the different medical experience and the presence of some blurred regions. Therefore, developing an effective and efficient method to automatically segment vessels from retinal images is very important for the diagnosis of related diseases.

\begin{figure}
\begin{center}
\includegraphics[width=3.5in]{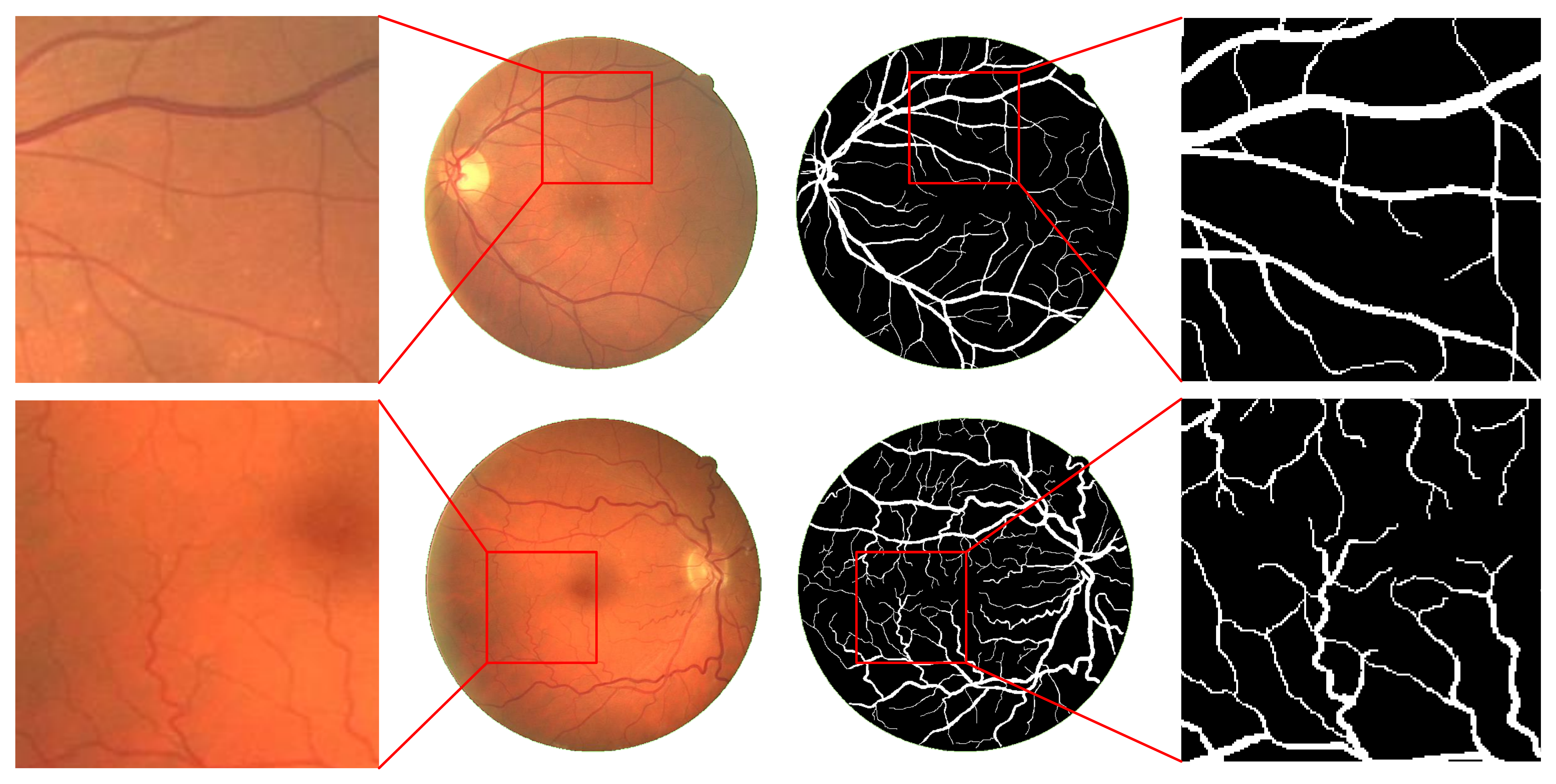}
\end{center}
   \caption{Examples of challenging areas in retinal blood vessel segmentation. First row: the blood vessels with various thicknesses and the corresponding ground truth; Second row: the blood vessels with blurry boundaries and the corresponding ground truth. Thickness inconsistencies and blurry boundaries make it challenging to accurately extract the vessels. (Best viewed in color.)}
\label{fig:Examples of our Retinal image segmentation results on two public benchmarks}
\end{figure}

Technically, retinal vessel segmentation requires one to accurately extract the pixel-wise locations of vessels regardless of different diameters, angles, and branching patterns. However, as is shown in Fig. \ref{fig:Examples of our Retinal image segmentation results on two public benchmarks}, segmentation of retinal vessel is challenging due to several reasons. First, the thicknesses of blood vessels vary largely in the retinal images. Such thickness inconsistencies make the thick vessels occupy more areas (i.e., pixels) than the thin vessels, which is referred to as multi-scale features learning problem \cite{sun2019neural, huang2020semantic}. Second, owing to the signs of ophthalmological diseases and unbalanced illuminations, a lot of blood vessel boundaries have low intensity contrast with nearby background \cite{zhao2018automatic}. Thus, most regions of thin and edge vessels are difficult to predict accurately. In addition, the amount of labeled data is insufficient because they require laborious annotations by domain experts.

\subsection{Related works}
\label{subsec:Related Works}

Recently, deep learning methods have been verified to achieve high accuracy for the retinal vessel segmentation and the diagnose of ophthalmological diseases using retinal images \cite{azzopardi2015trainable, xie2015holistically, zhou2017improving, wang2019blood}. Various neural network models are designed for learning the mapping from the fundus image to the labeled ground truth \cite{yu2020framework, ronneberger2015u, fu2016deepvessel, gu2019net, mou2019cs}. 
In particular, Yu {\em et al.} \cite{yu2020framework} use a deep neural network to segment the retinal blood vessels and then propose two algorithms for the hierarchical division of retinal vascular networks. To automatically detect hard exudate of diabetic retinopathy, Huang {\em et al.} \cite{huang2021new} propose a patch-based deep convolutional neural network, which effectively augments the training sample data and characterizes the retinal fundus images. Feng {\em et al.} \cite{feng2020ccnet} design a cross-connected convolutional neural network for retinal vessel segmentation by fusing multi-scale information. 
In these works, U-Net \cite{ronneberger2015u} is one of the most popular backbones for biomedical image segmentation, where the location information is refined by combining the high-resolution feature channels from the encoder with the de-convolutions features in the decoder structure. In order to improve the segmentation performance, a lot of works propose the extensions of the U-shaped network. For instance, Wu {\em et al.} \cite{wu2018multiscale} introduce max-pooling layers and up-sampling layers, and assemble them in two network followed network submodels. Wang {\em et al.} \cite{wang2019dual} design a dual encoding U-Net for retinal vessel segmentation, consisting of a spatial path and a context path, to preserve the spatial information and capture more semantic information respectively. {Yang {\em et al.} \cite{yang2021hybrid} propose a two-decoder multi-task segmentation U-Net to separately segment the thick and thin vessels, and then use a fusion network to fuse them.} These variants of U-shaped network further improve the segmentation results although they require large computational costs.

For the refined segmentation of retinal blood vessels, a key challenge is to detect vessels with various thicknesses. Some recent approaches on medical image segmentation use multi-scale networks to capture more context information \cite{hu2018retinal, fu2018joint, zhuang2018laddernet, zhang2019attention}.  
Hu {\em et al.} \cite{hu2018retinal} propose a multi-scale architecture to learn richer multi-scale vascular feature and design an improved cross-entropy loss function to pay more attention on hard examples. Similarly, Fu {\em et al.} \cite{fu2018joint} propose a multi-label deep network consisting of multi-scale input and side-output layers for joint optic disc and cup segmentation. However, the convolutions and pooling operations in fully convolutional networks (FCNs) may neglect the structural information during the encoding part. To address this drawback, Zhang {\em et al.} \cite{zhang2019attention} incorporate an attention guided filter into the network. Mou {\em et al.} \cite{mou2020dense} obtain the features in different layers and combine them through a multi-scale dice loss. Cherukuri {\em et al.} \cite{cherukuri2020deep} employ multi-scale representation filters to handle vessel thickness diversity. Moreover, Yan {\em et al.} \cite{yan2018joint} propose a novel segment-level loss that improves the segmentation results for thin vessels. Compared with single-scale models, multi-scale models can successfully predict more blood vessels but suffer from the problem of the classical trade-off: while emphasizing more on the thin vessels, it often leads to a slight decline in segmentation of the thick vessels.

\subsection{Motivation and contributions}
\label{subsec:Motivation and Contributions}

Whereas the above methods improve segmentation results by elaborate neural networks, they still have some limitations. First, when detecting blood vessels at multiple scales, most approaches design multi-scale modules (e.g., multiple encoders or decoders) and individually train them. {These modules compose an ensemble model, but lack sharing mechanisms to each other. Specifically, there is no interaction mechanism between the sub-models that are used to detect vessels with various thicknesses. For instance, in \cite{wu2018multiscale}, the patches are fed into two individual sub-models (up-pool network and the pool-up network) followed by an average operation on two final output features.} 
Meanwhile, compared with the single encoder-decoder architecture, multiple modules may exacerbate the problem of high computational complexity in retinal vessel segmentation task. In contrast to the obsessive interest on accuracy, the amelioration in efficiency cannot be neglected in medical image processing. Second, most deep learning-based methods simply couple the last layer with a loss function between the probability map and the manually annotated segmentation. There is no mechanism or regularization on hidden layers to guide an efficient learning for representation filters. Hence, they may learn inadequate even meaningless features. Furthermore, with the progress of training, the phenomenon of overfitting is easy to appear.

To address the above problems, we propose a novel deep neural NETwork based on Shared decoder and Pyramid-like loss (SPNet), for retinal blood vessel segmentation. Based on a U-shaped architecture, SPNet consists of an encoder and a novel decoder without extra network parameters. In the encoder, it uses convolution operations to extract deep features from the retinal images as original U-Net. In the decoder, it introduces a decoder-sharing mechanism capable of simultaneously learning multi-scale semantic information between the thick vessels and the thin vessels. Specifically, the high-level feature maps and the low-level ones are decoded at the same time in a sequence of shared decoder modules (SDMs). Compared with the simple combination of multi-scale decoders, SDMs help the model share knowledge between global and local structures, and then improve the semantic information extraction ability for thin and tiny vessels.

In addition, in order to decompose the semantic information that should be learned at various levels, we design a residual pyramid architecture. These residuals represent the potential errors when extracting contextual information among multi-scale features, especially the regions of capillaries and edge lines of vessels. Based on this residual pyramid architecture, we define a pyramid-like loss to guide the learning of representation filters in a multi-stage manner. Finally, by minimizing the global loss and the pyramid-like loss, our network obtains the probabilistic vessel map based on {\em error-compensation}.

We wish to emphasize the distinctive aspects of our method relative to some closely related works. 
{1) The way of multi-scale feature learning in our SPNet is different from some existing works.} In \cite{hu2018retinal}, Hu {\em et al.} propose a class-balanced cross-entropy loss function to ignore the loss of easy samples (i.e., well-predicted pixels) and enhance the learning of hard examples (i.e., poorly-predicted pixels). This loss is used in each side-output of the multi-scale architecture and relies on the parameter setting. 
{In this paper, we regard the decoding part of U-Net as a coarse-to-fine feature learning manner, where the spatial size of features is expanded by the up-sampling operation. The key in this process is to construct the new pixel points based on the coarse feature maps. To simulate the process of coarse-to-fine feature learning, in residual pyramid architecture, we up-sample the low-resolution ground truths (GTs) to the spatial size of the original GT. We refer to the difference between the up-sampled GTs and the original GT as the errors (or residuals).} 
This architecture of calculating the errors is parameter-free and enables the SPNet to learn the differences among multi-scale contextual information; 
{2) The proposed pyramid-like loss is different from the multi-scale loss in some representative methods.} The multi-scale loss \cite{fu2018joint, zhang2019attention, mou2020dense} measures the losses between outputs and labels at various resolutions, and then sums these losses as an objective function. These methods can overcome the problem of thickness inconsistency of blood vessels to a certain extent. 
{Moreover, based on the multi-scale loss, the network decodes the global features in a coarse-to-fine manner. The common semantic information in these multi-coarseness features is the background and main vessel trunks. Different from the multi-scale loss, the pyramid-like loss explicitly takes the losing pixels during the down-sampling process into account. In our SPNet, we construct the residuals to reflect the differences among multiple contextual information, which guides the decoder to tackle the challenges of thickness inconsistencies and blurry boundaries;} 
{3) The training process of SPNet is simpler than that of some works.} In \cite{cherukuri2020deep}, image patterns and noisy patches should be specially designed for learning knowledge on geometric feature and false positives in advance. Yan {\em et al.} \cite{yan2018joint} generate skeletons and divide them into several segments {before model training phase}. However, our SPNet requires no prior-processing step {to generate image patterns as \cite{cherukuri2020deep} or to divide the blood vessels into well-predicted and poorly-predicted parts as \cite{yan2018joint},} 
and directly carries out parameter optimization based on joint pyramid-like and global loss functions.

We summarize the contributions of this paper as follows.
\begin{itemize}
\item We propose a novel SPNet for retinal vessel segmentation, by employing the U-Net as the body structure without extra network parameters. The decoder-sharing mechanism and pyramid-like loss enable the network to focus on learning the residuals among semantic features at various levels.
\item We propose a set of SDMs to learn semantic information at diverse scales in the feature maps. They not only help the network share knowledge among multi-scale contexts, but also potentially prevent overfitting.
\item We propose a residual pyramid that captures the information lost in the multi-scale features decoding process. To the best of our knowledge, this paper is the first to specially design an approach with residual pyramid architecture for retinal blood vessel segmentation. We also provide a pyramid-like loss function to integrate the SDMs and the defined residual pyramid.
\item We validate the effectiveness of our SPNet on three popular benchmark datasets, i.e., DRIVE dataset \cite{staal2004ridge}, STARE dataset \cite{hoover2000locating}, and CHASE\_DB1 dataset \cite{owen2009measuring}. The experimental results show that the proposed method outperforms the original U-Net and the state-of-the-art methods.
\end{itemize}

The rest of this paper is organized as follows. In Section \ref{sec:Shared decoding U-Net with pyramid-like loss}, the proposed SPNet is presented. In Section \ref{sec:Experiments}, the proposed method is evaluated by experiments on three public datasets. Analysis and discussion on our work are described in Section \ref{sec:Discussion}. Conclusions and future works are given in Section \ref{sec:Conclusion}.

\section{Methods}
\label{sec:Shared decoding U-Net with pyramid-like loss}

In this section, we first present backgrounds with mathematical notations. Then we introduce the decoder-sharing mechanism to capture the multi-scale semantic information, and describe the residual pyramid which defines the missing information during the down-sampling process. Finally, we present how to integrate them into our SPNet.

\subsection{Background and notation}
\label{subsec:Background and Notation}

Let us begin with the definitions of terminologies. Denote the input image as $I \in \mathbb{R}^{H \times W}$, where $H$ and $W$ are the height and width respectively. Denote the final output probability map as $O_{0} \in \mathbb{R}^{H \times W}$ and the ground truth as $G_{0} \in \{0, 1\}^{H \times W}$. The proposed architecture consists of $2L \ (L \in \mathbb{N}^+)$ layers, where the first $L$ layers are the encoding part and the rest are the decoding part. As the U-Net \cite{ronneberger2015u}, our SPNet has double-convolution layer (in the first $2L-1$ layers), which contains two groups of convolution and rectified linear unit (ReLU) operation. Mathematically, the mapping function in the layer $l$ is defined as follows,
\begin{equation}
\label{equ:double-convolution}
f_{l}(\cdot) \equiv \max(c_{l,2}(\max(c_{l,1}(\cdot), 0)), 0), ~ l=1,2,\cdots,2L-1,
\end{equation}
where $c_{l,j}(\cdot)$ is a $3 \times 3$ convolution operation. Among these double-convolution layers, down-sampling or up-sampling is implemented to extract contextual information at various levels. To be specific, the output of $l$-th layer in encoding part is
\begin{equation}
E_l = f_{l}(d_l(E_{l-1})), ~ l=1,2,\cdots,L,
\end{equation}
where $d_l(\cdot)$ is a down-sampling of $(\cdot)$ and $d_1(E_{0})=I$. Through the above operations, the level of feature map raises along with the reduction of resolution. However, some semantic information loses due to the down-sampling. To overcome this disadvantage, the skip connections are adopted between the encoding and decoding parts. Correspondingly, the output of $l$-th layer in decoding part is
\begin{equation}
D_l = f_{L+l}[u_l(D_{l-1}), E_{L-l}], ~ l=1,2,\cdots,L-1,
\end{equation}
where $u_l(\cdot)$ is an up-sampling of $(\cdot)$ and $D_{0}=E_{L}$. In the last layer, features are transformed to the probability map $O_{0}$ using a $1 \times 1$ convolution operation followed by a sigmoid function, which can be formulated as:
\begin{equation}
\label{equ:last-convolution}
O_{0} = f_{2L}(D_{L-1}).
\end{equation}
Finally, the network parameters are learned by minimizing the loss function. As the majority of recent works \cite{wu2018multiscale, yan2018joint}, the proposed network generates a probability map $O_{0}$, which is then transformed to a hard segmentation result using a threshold of 0.5.

\begin{figure*}
\begin{center}
\includegraphics[width=1\textwidth,trim=0 0 0 0,clip]{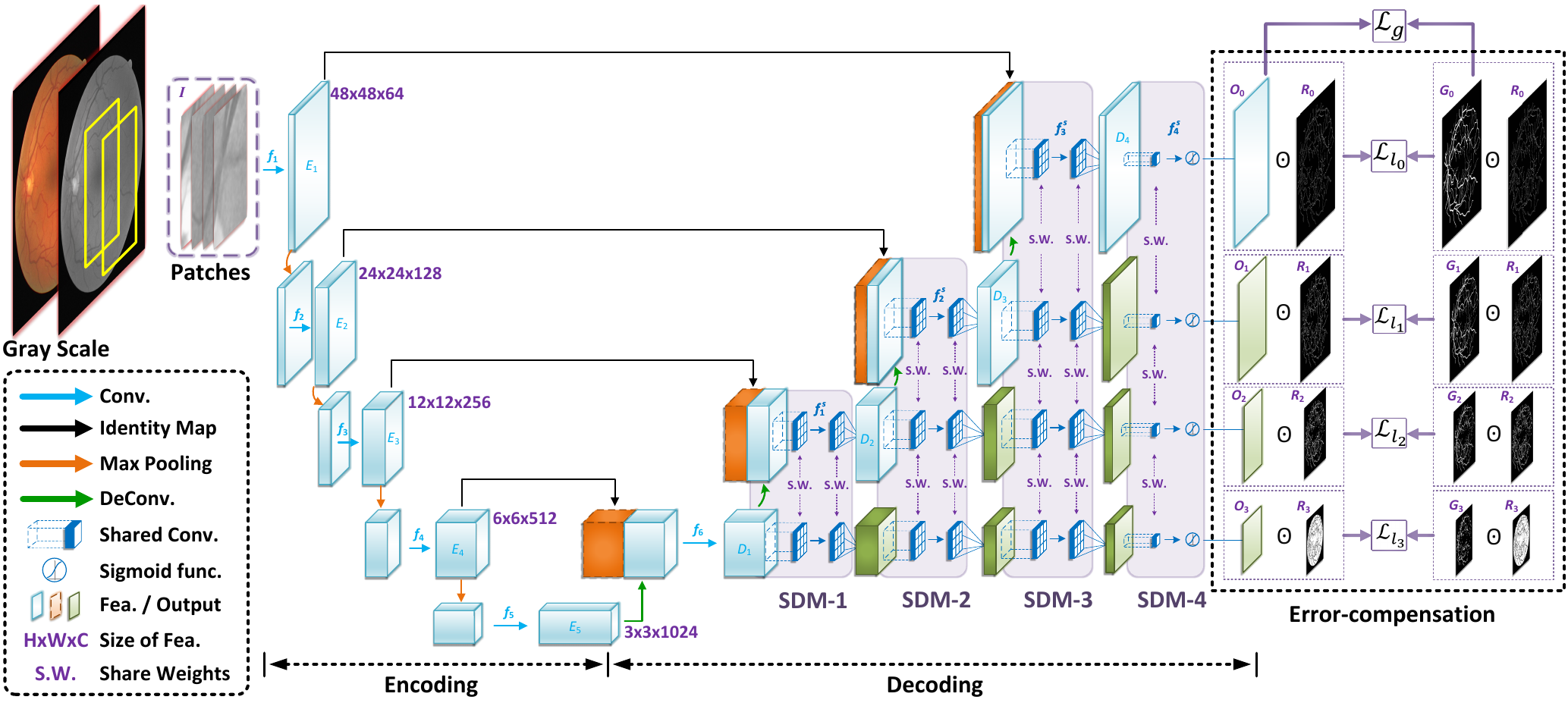}
\end{center}
   \caption{{Overall architecture of the proposed SPNet. It employs the U-Net as the body structure. In the encoding part, five double-convolution layers are used. In the decoding part, without introducing extra network parameter, the designed SDMs are carried out to obtain multiple final probability maps $O_{k}$s. Finally, the network obtains the refined segmentation result by minimizing a global loss $\mathcal{L}_{g}$ and the pyramid-like loss, i.e., the fusion of the local losses $\mathcal{L}_{l_{k}}$s. (Best viewed in color.)}}
\label{fig:Overall architecture of the SPNet}
\end{figure*}

An overview of the proposed network is shown in Fig. \ref{fig:Overall architecture of the SPNet} where the number of layers is $2L = 10$. Employing the U-Net as the body structure, SPNet has an encoder-decoder architecture. In the encoding part, it has five double-convolution layers. In the decoding part, it consists of a double-convolution layer and four SDMs. Finally, a pyramid-like loss based on the residual pyramid is proposed, which helps in compensating errors on the capillaries and the vessel contours. Since the deep features obtained from the encoder are combined with the multi-scale features in the decoder by skip connections, the modification on decoder will indirectly improve the learning of encoder.

\subsection{The decoder-sharing mechanism}
\label{subsec:Decoder-sharing Mechanism}

In the task of vessel segmentation, a key challenge is to predict vessels with various thicknesses. Main vessel trunks can often be successfully extracted, since they occupy the vast majority of areas and show a striking contrast with the background. However, as is shown in Fig. \ref{fig:Examples of our Retinal image segmentation results on two public benchmarks}, the blurry boundaries and variable orientations hamper the segmentation for the thin and tiny vessels. Some recent works \cite{fu2018joint, zhang2019attention, mou2020dense} employ extra convolution operations on the features in the decoding part, and then obtain multiple outputs. As an inherent drawback among most convolutional neural networks, the computational complexity looms large in the task of segmentation for the improvements in efficiency. {Besides, from the experiment results, we observe that simple decoding operation, i.e., a $1 \times 1$ convolution, on the high-level features cannot obtain enough semantic information on the vessel structure.} Instead of extracting features for each deep layer using respective filters, SDMs are designed to learn multi-scale contextual information simultaneously in a decoder. Compared with the original decoder in U-Net, SDMs can bring great improvement for retinal vessel segmentation without introducing any parameters.

Since the first layer in the decoding part is a double-convolution function of the deepest features (concatenation of $u_{L+1}(E_L)$ and $E_{L-1}$), the SDMs begin in the second decoding layer and then the number of SDMs is $L-1$. Denote SDM-$i$ ($i \in \left \{ 1, 2, \cdots, L-1 \right \}$) as the $i$-th SDM, which has a $2 \times 2$ de-convolution operation $u_{i}(\cdot)$ and a shared convolution $f_{i}^{s}(\cdot)$ defined in \eqref{equ:double-convolution} or \eqref{equ:last-convolution} (for notational ease, we use $f_{i}^{s}(\cdot) \equiv f_{L+1+i}(\cdot)$ and the superscript ``s'' for sharing the weights). Mathematically, if $i \leq L-2$, in SDM-$i$,
\begin{align}
D_{i+1}&=f_{i}^{s} [u_{i}(D_{i}), E_{L-i-1}],\label{equ:D_i+1}\\
F_{i+1,j+1}&=f_{i}^{s} (F_{i,j}), ~ j=0, 1, \cdots, i-1,\label{equ:Fi+1_j+1}
\end{align}
where $D_{i+1}$ and $F_{i+1,j+1}$s are the obtained feature maps, and $F_{i,0}=D_{i}$. If $i = L-1$, SDM-$i$ generates multiple probability maps as:
\begin{equation}
\label{equ:O_j}
O_{j} =
\begin{cases}
f_{L-1}^{s} (D_{L-1}),&\ j = 0, \\
f_{L-1}^{s} (F_{L-1,j}),&\ j=1, \cdots, L-2.
\end{cases}
\end{equation}
Finally, the progression $\{$SDM-$i \}_{i=1}^{L-1}$ forms a sequence of weight-sharing decoder modules. 
{As is shown in Fig. \ref{fig:Overall architecture of the SPNet}, the parameters are shared among the convolution operations in different scales in the same SDM. Note that the parameters are shared neither between the two convolution operations in a double-convolution layer (i.e, $c_{l,1}$ and $c_{l.2}$ in \eqref{equ:double-convolution}), nor among the convolution operations in the decoder in the same scale.}

Through the iterations from \eqref{equ:D_i+1} and \eqref{equ:Fi+1_j+1} to \eqref{equ:O_j}, we can make several observations. On one hand, {the decoding part to obtain multi-scale features $D_{L-1}$ and $F_{L-1,j}$s has the same number of convolution operations as the encoding part, whereas in the model with 1x1 convolution side-output operations, the number of convolution operations in the decoding part is less than that in encoding part. Comparing with this 1x1 convolution operation, our SPNet has a certain type of symmetry on the number of convolution operations. The experiment results on the feature maps and final segmentations validate that this symmetry helps us to extract more contextual information on the vessel structure.} On the other hand, the decoders for multi-scale features $D_{i}$ and $F_{i,j}$s share the convolution weights. This has a similar manner with the powerful operation, Atrous Spatial Pyramid Pooling module \cite{chen2017deeplab, chen2018encoder}, which employs parallel atrous convolution with different rates to encode multi-scale semantic information. In our method, the proposed SDMs simultaneously extract features at multiple scales. They help the model share knowledge learned at different resolutions and then improve the segmentation results for the challenging areas.

The advantages of the SDM are three-fold.
\begin{itemize}
\item First, compared with the decoding operations in U-shaped architecture, it widens the width of convolutional neural network without introducing any extra parameters. {Specifically, in the raw decoder path of U-Net, there is one double-convolution operation in each layer. In the decoder path of our SPNet, there are multiple double-convolution operations (with shared weights) in each layer. From the perspective of final prediction, U-Net contains only one probability map, whereas our SPNet has multi-scale probability maps.}
\item Second, it simultaneously learns multi-scale semantic information in the feature maps. In each SDM, de-convolution and concatenation operations introduce more semantic information to high-resolution low-level features. The weight-sharing mapping function decodes $D_{i}$ and $F_{i,j}$s individually while encouraging that the decoder shares knowledge among them. This not only proactively learns the multi-scale contextual information but also potentially prevents overfitting. {The idea of weight-sharing among multi-scale features is also explored in \cite{li2019scale}, where Li {\em et al.} use the same parameters for multiple scale ranges under different receptive, and the experiment results validate the effectiveness of weight-sharing mechanism.}
\item Third, it provides a new approach to visualize the feature maps in the decoding part. In \cite{fu2018joint}, an up-sampling and a $1 \times 1$ convolution are directly adopted to obtain the multi-label prediction maps, which may neglect some high-level semantic information from the encoding phase. However, SDM contains appropriate operations for each single-scale output in the decoding part.
\end{itemize}

\subsection{The residual pyramid architecture}
\label{subsec:Residual Pyramid Architecture}

Next, we design a residual pyramid to decompose the errors when refining the retinal vessel segmentation. This architecture guides our model to learn the differences among multi-scale contextual information. We will integrate it into the final network with SDMs in Section \ref{subsec:Objective Function}.

\begin{figure}[t]
\begin{center}
\includegraphics[width=3.5in]{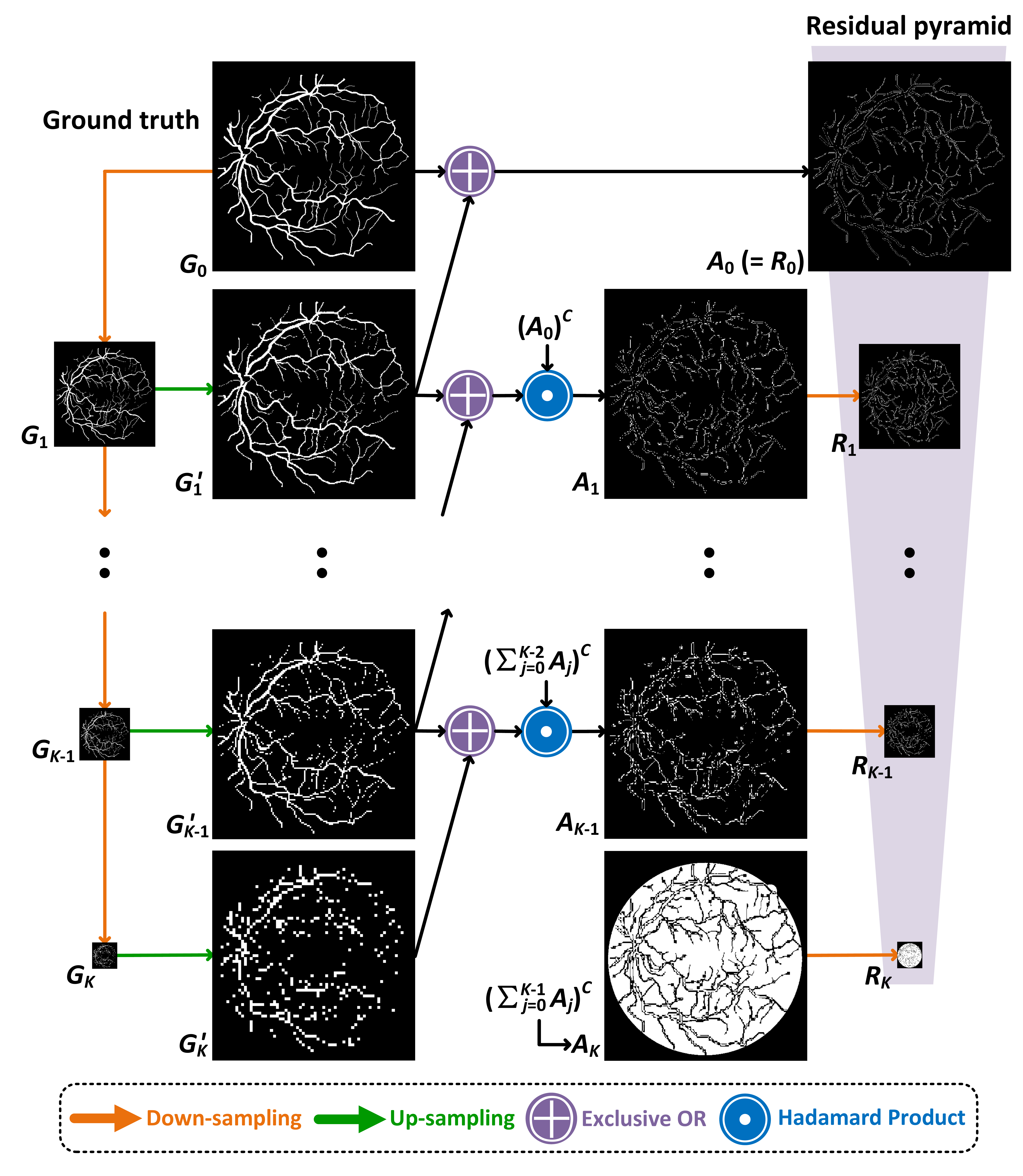}
\end{center}
   \caption{The residual pyramid architecture. {All images are in binary (1: white pixel; 0: black pixel).} First, the multi-scale ground truth $G_{k}$s are expanded to the same spatial size (i.e., from $G_{k}$ to $G_{k}'$), followed by the exclusive OR operations to capture the differences. Second, the Hadamard product operations of the differences and the complementary sets $(\sum\nolimits_{j=0}^{k} A_{j})^{C}$ are implemented to ensure that error for each pixel is compensated in the higher level of pyramid. Finally, $A_{k}$s are scaled-down by the factor of $2^{k}$ to obtain residual pyramid $R_{k}$s. (Best viewed in color.)}
\label{fig:Residual pyramid}
\end{figure}

As is shown in Fig. \ref{fig:Residual pyramid}, first the raw segmentation ground truth $G_{0}$ is shrunk by $2^k$-fold to obtain multi-scale labels $G_{k}$s. Second, $G_{k}$s are expanded to same spatial size as $G_{0}$, denoted by $G_{k}'$s. Third, an exclusive OR operation is implemented between any two successive members of $G_{k}'$s. We can formulate the procedure as
\begin{equation}
\label{equ:calculate for A_k}
A_{k}=
\begin{cases}
G_{k}' \oplus G_{k+1}',&\ k = 0, \\
( G_{k}' \oplus G_{k+1}' ) \odot (\sum\nolimits_{j=0}^{k-1} A_{j})^{C},&\ k = 1, \cdots, K-1 \\
(\sum\nolimits_{j=0}^{k-1} A_{j})^{C},&\ k = K,
\end{cases}
\end{equation}
where $G_{0}' = G_{0}$, $\oplus$ and $\odot$ denote the exclusive OR operation and the Hadamard product respectively, and $(\cdot)^{C}$ is the element-wise complementary set of the binary image. In \eqref{equ:calculate for A_k}, we use the exclusive OR operation (simply computed by the absolute value $\left | G_{k}' - G_{k+1}'  \right |$) instead of $(G_{k}' - G_{k+1}')$. This choice enables us to integrate it into the pyramid-like loss function. Moreover, the $A_{k}$s capture the difference between the adjacent levels and filter out {the pixels with binary value of 1} in the higher level of pyramid. In such an architecture, the numerical optimization considers each pixel during the error-compensation phase. Finally, the sizes of $A_{k}$s are scaled-down by the factor of $2^{k}$, and then residual pyramid $R_{k}$s are obtained.

{In practice, we down-sample and up-sample the ground truths using the nearest-neighbor interpolation. The motivation of down-sampling and up-sampling the ground truth labels is to simulate the process of losing information (as the encoding part of U-Net) and capture the loss information (to guide the learning of the decoding part). Specifically, by down-sampling $G_{0}$, we obtain multi-scale labels $G_{k}$s, which can be viewed as the labels corresponding to the multiple side-output prediction maps in the decoder. However, we do not directly use $G_{k}$s to define a multi-scale loss, since the common semantic information among them mainly include the background and main vessel trunks. We expect that the decoding part focuses on learning the differences among multi-scale labels, i.e., the contours and small vessels. By up-sampling $G_{k}$s to the same spatial size as $G_{0}$, we obtain the labels with varying degrees of coarseness $G_{k}' $s. To capture the differences among multi-coarseness labels, we implement an exclusive OR operation on the adjacent ones. After calculating the Hadamard product operation of the differences and the complementary set, each pixel in the whole image is considered only once, i.e., $\sum_{k=1}^K = A_k = M$, where $M$ is the field of view. Finally, we down-sample $A_k$s so as to match the size of side-output prediction maps in the decoder.}

It can be observed in Fig. \ref{fig:Residual pyramid} that the proposed residual pyramid architecture decomposes the pixels in a ground truth into local features. {These local features will be integrated into a pyramid-like loss (defined in Section \ref{subsec:Objective Function}) on the decoding part of our SPNet. Note that we also use the global loss between the ground truth $G_0$ and the final prediction map $O_0$. Hence, in the training process of SPNet, the network parameters are updated based on two loss terms, i.e., the global loss function and the pyramid-like loss function.} In the proposed residual pyramid architecture, $R_K$ highlights the regions of background and main vessel trunks, whereas the rest focus on the contours and small vessels. Such an architecture possesses two properties as follows. First, it tackles the challenges of thickness inconsistencies in a multi-stage learning mechanism (each hidden layer concentrates on a residual in the pyramid). Second, the representation filters in each hidden layer are the results of optimization based on training data coupled with the corresponding residual pyramid, which has an adaptability to dataset and can directly improve segmentation performance. 

Some existing works, e.g., \cite{ghiasi2016laplacian} and \cite{zhao2018pyramid}, use the pyramid framework to predict probability map in a coarse-to-fine manner. This type of refinement process obtains the segmentation results for approaching the multi-scale labels (from $G_{K}$ to $G_{0}$). 
However, this procedure introduces redundant information in an end-to-end neural network. Furthermore, during the decoding phase, the segmentation process may ignore the high-level semantic information mapping in deep encoder layers. In the proposed residual pyramid, the exclusive OR operations between the $G_{k}'$s are employed so that $R_{k}$s can capture the differences of semantic labels among different resolutions.

\subsection{Objective function}
\label{subsec:Objective Function}

In our method, the loss function contains two terms, a global loss function $\mathcal{L}_{g}$ and the pyramid-like loss function $\mathcal{L}_{l_{k}}$s. The global loss considers the pixel-wise difference between the probability map and the ground truth. Here, we use multi-scale Dice loss function \cite{sudre2017generalised}, i.e.,
\begin{equation}
\label{equ:L_dice}
\mathcal{L}_{g} = 1 - \frac{2 \mathbf{e}^\top(O_{0} \odot G_{0})\mathbf{e} + \epsilon}{\left \| O_{0} \right \|_\mathrm{F}^2 + \left \| G_{0} \right \|_\mathrm{F}^2 + \epsilon},
\end{equation}
where $\mathbf{e} = {[1, \cdots, 1]}^\top$, $\left \| \cdot \right \|_\mathrm{F}$ denotes the Frobenius norm, and {$\epsilon$ is a smoothing factor to prevent the denominator becoming zero when $O_{0}$ and $G_{0}$ are empty, i.e., the patch is a pure background without any vessel pixels.} $\mathcal{L}_{g}$ takes into consideration all pixels equally, including the background and the blood vessels. Unbalanced proportions of regions in the retinal images may lead the model to excessively learn the semantic information on the background and main vessel trunks, while neglecting the contours and small vessels.

In order to strengthen characterization on these challenging areas (the capillaries and the edges of blood vessels), we design a pyramid-like loss function, which consists of several local modules. We employ the residual pyramid (i.e., $R_{k}$s) to define the local loss modules. Such a residual pyramid, like {\em filter}, helps the decoder to ignore the pixels that have been classified correctly, and focus on those undetermined and misclassified regions. The loss terms are defined as follows:
\begin{equation}
\mathcal{L}_{l_{k}} = \mathcal{L}_{CE}(G_{k} \odot R_{k}, O_{k} \odot R_{k}), \ k = 0, 1, \cdots, K,
\end{equation}
where $\mathcal{L}_{CE}(\cdot)$ denotes the cross-entropy loss function. $\mathcal{L}_{l_{K}}$ is the loss module focusing on the regions of the most background and the centerline of main vessel trunks, whereas $\mathcal{L}_{l_{k}}$s ($k = 0, 1, \cdots, K-1$) focus on the regions that should be compensated in the decoder, especially the capillaries and the edge lines of blood vessels.

The final loss function in our SPNet is:
\begin{equation}
\label{equ:final loss function in our SPNet}
\mathcal{L}_{t} = \underbrace{\mathcal{L}_{g}}_{\text{Global loss}} + \underbrace{\sum\nolimits_{k=0}^{K} \lambda_{k} \mathcal{L}_{l_{k}}}_{\text{Pyramid-like loss}},
\end{equation}
where $\lambda_{k}$s are the trade-off parameters. The gradients of $\mathcal{L}_{t}$ with respect to $O_{k}$s are expressed as:
\begin{align}
\begin{split}
\frac{\partial \mathcal{L}_{t}}{\partial O_{0}} = &- \lambda_{0} G_{0} \odot R_{0} \oslash O_{0} - \frac{2 G_{0}}{\left \| O_{0} \right \|_\mathrm{F}^2 + \left \| G_{0} \right \|_\mathrm{F}^2 + \epsilon} \\
&+ \frac{2 [2 \mathrm{e}^\top(O_{0} \odot G_{0})\mathrm{e} + \epsilon] O_{0}}{(\left \| O_{0} \right \|_\mathrm{F}^2 + \left \| G_{0} \right \|_\mathrm{F}^2 + \epsilon)^{2}},
\end{split}
\\
\frac{\partial \mathcal{L}_{t}}{\partial O_{k}} =& - \lambda_{k} G_{k} \odot R_{k} \oslash O_{k}, \ k = 1, 2, \cdots, K.
\end{align}
where $\oslash$ is the Hadamard {division operation}. {For conciseness, denote $s(\cdot | \phi)$ as the SPNet with network parameter $\phi$, i.e, ${\{O_k\}}_{k=0}^{K} = s(I | \phi)$. During each iteration in the training phase, we first calculate the error function $\mathcal{L}_{t}$ by \eqref{equ:final loss function in our SPNet}. Then, we calculate the gradient of $\mathcal{L}_{t}$ with respect to $\phi$ as $\frac{\partial \mathcal{L}_{t}}{\partial \phi} = \sum_{k=0}^K \frac{\partial \mathcal{L}_{t}}{\partial O_k} \frac{\partial O_{k}}{\partial \phi}$. Finally, we update the network parameter via $\phi' = \phi - \eta \frac{\partial \mathcal{L}_{t}}{\partial \phi}$ with learning rate $\eta$.} 
Note that the parameters in SDMs depend directly on $\mathcal{L}_{l_{k}}$, and hence they are updated in gradient computation. The constraints based on the residuals from low-resolution label will influence all shared convolutional filters $f_i^s(\cdot)$, which is the same as multi-scale loss \cite{fu2018joint, zhang2019attention, mou2020dense}. However, our SPNet decomposes the differences of multiple resolutions, and then focuses on the learning of region within each $R_k$ in each stage of the decoder. This shows that the decoder in our method has interpretability, namely compensating errors. Also, our pyramid-like loss function has a significant difference with \cite{zhao2018pyramid}. To be specific, the Pyramid-Based FCNs in \cite{zhao2018pyramid} defines a residual mask to make the lower level FCN pay less attention to the pixel when the upper level FCN achieves good segmentation and more attention to bad segmentation. This procedure calculates the losses on pixels existing in all scales for $K$ times and inevitably leads to overfitting on those easily segmented pixels. On the contrary, our SPNet contains one FCN, and the local loss on each pixel in the pyramid-like loss function is considered only once. Taking the global loss and the pyramid-like loss into account, SPNet can successfully predict not only the main vessel trunks but also the capillaries and the edge lines of blood vessels.

\section{Experiments and analysis}
\label{sec:Experiments}

To evaluate our method, we carry out comprehensive experiments on DRIVE dataset\footnote{http://www.isi.uu.nl/Research/Databases/DRIVE/.} \cite{staal2004ridge}, STARE dataset\footnote {http://cecas.clemson.edu/~ahoover/stare/.} \cite{hoover2000locating}, and CHASE\_DB1 dataset\footnote {https://blogs.kingston.ac.uk/retinal/chasedb1/.} \cite{owen2009measuring}. We first introduce the datasets and implementation details. Second, we perform a series of ablation experiments. Finally, we report the segmentation results on the three datasets.

\subsection{Datasets}
\label{subsec:Datasets}

The DRIVE dataset contains 40 color retina images, with a $45^{\circ}$ field of view at $565 \times 584$ pixels. For each image, the manual vessel segmentation is provided. Following the standard partition, 20 images are used for training and the other 20 for testing.

The STARE dataset includes 20 color retina images, with a $35^{\circ}$ field of view at $700 \times 605$ pixels. The hand labeled segmentation result is provided for each color image. Since there is not a standard partition given, we follow the leave-one-out validation strategy \cite{yu2020framework}. That is, 19 images are used for training and the remaining 1 for testing, which are repeated 20 times.

The CHASE\_DB1 dataset consists of 28 color retina images, where the size of each is $999 \times 960$. The segmentation ground truth is also provided for each image in CHASE\_DB1. Normally, a 4-fold cross-validation strategy is used on this dataset. In each fold, 21 images are used for training and the other 7 for testing.

\subsection{Implementation details}
\label{subsec:Implementation Details}

Since the benchmark datasets provide only a small number of training images, we implement data augmentation strategy as follows. For each training image in DRIVE dataset, we randomly pick a pixel as the center of a patch with size $48 \times 48$ and then extract 9,500 patches, from which we get a total of 190,000 tiles. For STARE dataset, we randomly extract 19,000 patches from each retinal image and then obtain a total of 361,000 tiles. For CHASE\_DB1 dataset, we randomly extract 38,000 patches allowing overlapping from each image and then obtain a total of 798,000 tiles.

During training, 70\% of the samples (i.e., patches) are used and the remaining 30\% are taken as validation. {To avoid the problem of exactly overlaps in the training samples and the validation samples, we use a block-based data splitting strategy as follows. In a randomly selected training image, we randomly generate a block and take all patches whose center points are inside this block into the validation set. We repeat this block generating operation until the sample size of the validation set reaches 30\% of the total sample size. Note that only those patches near the side of the block (strictly speaking, 24 pixels near the side) contain partial region-overlaps in the training and the validation. Overall, most patches in the training set and the validation set do not overlap.}

During testing, the overlap-tile strategy \cite{ronneberger2015u} is exploited. First, each test image is sliced into patches with size $32 \times 32$. Then, for each patch, $8$ pixels nearby the edges are used to extend its size; thus it has the same resolution as the training samples. Such a strategy makes it better to extrapolate the semantic labels with rich contexts, especially for the border region. Finally, the segmented patches are stitched in an orderly fashion to obtain the complete prediction map.

{In the training phase, we calculate the validation loss over five batches of data every ten training iterations, and choose the best iteration whose validation loss function value is the smallest. The hyper-parameters $\lambda_k$s are tuned based on 4-fold cross-validation on the training data (based on the separation of patients) using a grid search: $\lambda_0 \in \{10,1,0.5\}$, $\lambda_1 \in \{5,0.5,0.25\}$, $\lambda_2 \in \{2.5,0.25,0.125\}$, $\lambda_3 \in \{1.25,0.125,0.0625\}$. We tune the learning rate using a grid search within the set $\{1e{\text -}1,1e{\text -}2,1e{\text -}3,1e{\text -}4,1e{\text -}5\}$. For each value in the set, a line plot of loss over training iteration is plotted. We select the value when the reduction in loss is neither very slow (corresponding to a too small learning rate) nor very fast even oscillatory (corresponding to a too large learning rate). Since we terminate the training process when the validation loss starts increasing, the number of epochs is not so significant. We set the number of epochs to be 20 in our experiments. As for the batch size, we let it be in power of 2 and as large as possible to fit into one GPU memory.}

All the experiments are implemented based on PyTorch 1.2.0. Adam optimizer with initial learning rate $\eta_0 = 0.001$, $\beta_{1} = 0.5$, and $\beta_{2} = 0.999$ and batch size of 256 are applied over 20 epochs. For the sake of stable training, we let the learning rate decrease as $\eta_t = \eta_0 (1 - t/T)^{0.9}$ at the $t$-th iteration, where $T$ is the total number of iterations. In addition, batch normalization \cite{pmlr-v37-ioffe15} is carried out to prevent overfitting. Empirically, state-of-the-art performance can be achieved without extensive hyper-parameter tuning. {On all benchmark datasets, the smoothing factor $\epsilon$ is set as 1 in the global loss function} and the trade-off parameters in the pyramid-like loss functions are set as $\lambda_{0}=1$, $\lambda_{1}=0.5$, $\lambda_{2}=0.25$, and $\lambda_{3}=0.125$.

{The measurements of segmentation results are examined on the whole image level.} Four evaluation metrics, including sensitivity (Sen), specificity (Spe), accuracy (Acc) and AUC are employed. Pixels in the prediction map are measured by four indexes, i.e., True Positive ($TP$), True Negative ($TN$), False Positive ($FP$), and False Negative ($FN$). Sensitivity and specificity present the detailed segmentation performance on the vessels and the background respectively, which are defined as
\begin{align}
{\rm Sen}&=TP/(TP+FN),\\
{\rm Spe}&=TN/(TN+FP).
\end{align}
The accuracy delivers the overall segmentation performance:
\begin{equation}
{\rm Acc}=(TP+TN)/(TP+TN+FP+FN),
\end{equation}
and the AUC reports the area under the receiver operating characteristic curve.

\subsection{Ablation study}
\label{subsec:Ablation Study on DRIVE}

The decoder-sharing mechanism and the pyramid-like loss function are evaluated here. For simplicity, we implement models with different settings on DRIVE dataset and compare them with the baseline U-Net \cite{ronneberger2015u} {and some variants with the use of deep supervision. To perform the ablation experiments in the same conditions, we assign all models in the ablation studies with the same initial weights of layers. As for the training data, we use a random seed to keep the extracted patches and the data splitting (training versus validation) the same among all ablation studies. In addition, we do not use the parallel workers here.}

\begin{table}[!t]
\caption{{Ablation Study on the DRIVE dataset. The higher the metric values, the better the method. The values with bold symbol and underline present the best results and the second best one respectively. ``1x1conv'' and ``ML'' mean using $1 \times 1$ convolution operation and multi-scale loss \cite{zhang2019attention} respectively. ``SPNet (up)'' uses another strategy to compute the pyramid-like loss, i.e., up-sampling the feature maps to the resolution of the ground truth.}}
\label{tab:Ablation Study on DRIVE dataset.}
\normalsize
\renewcommand\tabcolsep{2.0pt}
\begin{center}
\begin{tabular}{c|cccccccc|c|c|c|c}
\toprule
Setting & SDMs & $\mathcal{L}_{g}$ & $\mathcal{L}_{l_{3}}$ & $\mathcal{L}_{l_{2}}$ & $\mathcal{L}_{l_{1}}$ & $\mathcal{L}_{l_{0}}$ & 1x1conv & ML & Acc & Sen & Spe & AUC \\
\midrule
Backbone &         & $\surd$ &         &         &         &         & && 0.9501             & 0.7356             & 0.9602             & 0.9641 \\
\cline{1-13}
S-1    &         & $\surd$ & $\surd$ & $\surd$ & $\surd$ & $\surd$ & && \underline{0.9672}    & 0.7776             & \textbf{0.9855}    & 0.9738 \\
\cline{1-13}
S-2    & $\surd$ & $\surd$ & $\surd$ &         &         &         & && 0.9617             & 0.7949             & 0.9779             & 0.9750 \\
S-3    & $\surd$ & $\surd$ & $\surd$ & $\surd$ &         &         & && 0.9640             & 0.8028             & 0.9796             & 0.9766 \\
S-4    & $\surd$ & $\surd$ & $\surd$ & $\surd$ & $\surd$ &         & && 0.9646             & 0.8193             & 0.9787             & 0.9763 \\
\textbf{SPNet} & $\surd$ & $\surd$ & $\surd$ & $\surd$ & $\surd$ & $\surd$ & && 0.9664 & \underline{0.8243}    & 0.9802 & \underline{0.9828} \\
S-5    & $\surd$ & $\surd$ &         & $\surd$ & $\surd$ & $\surd$ & && 0.9614             & 0.8222 & 0.9749             & 0.9748 \\
S-6    & $\surd$ & $\surd$ &         &         & $\surd$ & $\surd$ & && 0.9646             & 0.8139             & 0.9792             & 0.9760 \\
S-7    & $\surd$ & $\surd$ &         &         &         & $\surd$ & && 0.9621             & 0.8029             & 0.9776             & 0.9750 \\
\cline{1-13}
S-8 &         & $\surd$ &         &         &         &         & $\surd$ & $\surd$ & \underline{0.9672} &  0.7820 &  \underline{0.9851} &  0.9684 \\
S-9 &   $\surd$      & $\surd$ &         &         &         &         &  & $\surd$ & 0.9667 &  0.7896 &  0.9838 &  0.9771 \\
S-10 &         & $\surd$ & $\surd$ & $\surd$ & $\surd$ & $\surd$ & $\surd$ & & \textbf{0.9683} &  0.8205 &  0.9827 &  \textbf{0.9841} \\
\cline{1-13}
SPNet (up) & $\surd$ & $\surd$ & $\surd$ & $\surd$ & $\surd$ & $\surd$ & && 0.9612 &  \textbf{0.8363} &  0.9733 &  0.9776 \\
\bottomrule
\end{tabular}
\end{center}
\end{table}

{\em 1) Ablation study for the decoder-sharing mechanism:} To justify the effectiveness of SDMs, we replace them with multiple independent decoders, i.e., the functions $f_{k}^{s}$ ($k=1, 2, 3$) are still used but those parameters are not shared between the features with different sizes. The results are shown in Table \ref{tab:Ablation Study on DRIVE dataset.} (line of S-1). Although S-1 has the second highest accuracy and the highest specificity, it has much lower sensitivity and AUC, compared with those models with SDMs. This shows that the setting with multiple independent decoders is prone to overfitting on the background. In addition, the number of learnable parameters of S-1 increases by about $9.63\%$ than that of SPNet. These prove the superiority of our decoder-sharing mechanism.

{\em 2) Ablation study for the pyramid-like objective function:} Compared with the backbone U-Net, S-1 obtains improvements on all metrics. This verifies the beneficial effects of the pyramid-like loss function based on the residual pyramid. Next, we study the effect of the items in the pyramid-like loss function. Note that it consists of four local losses $\mathcal{L}_{l_{k}}$s. $\mathcal{L}_{l_{3}}$ stands for the loss module focusing on the regions of background and the centerline of main vessel trunks, whereas $\mathcal{L}_{l_{2}}$, $\mathcal{L}_{l_{1}}$ and $\mathcal{L}_{l_{0}}$ are those focusing on the regions of the capillaries and the contours, which should be compensated in the decoder.

As is shown in Table \ref{tab:Ablation Study on DRIVE dataset.}, the ablation study is designed in dual orders. Lines of S-2 to SPNet form a gradually increasing of loss functions in the decoding part. Compared with the backbone (only $\mathcal{L}_{g}$ is used), employing $\mathcal{L}_{l_{3}}$ yields extremely high specificity and sensitivity, which shows that excessively learning on the background and the centerline of main vessel trunks is carried out. Along with the joining of $\mathcal{L}_{l_{2}}$, $\mathcal{L}_{l_{1}}$ and $\mathcal{L}_{l_{0}}$, more and more errors on the capillaries and the edge lines of blood are compensated. {From the perspective of model training, the increasing of metrics from lines S-2 to SPNet can be explained by the fact that the network has more gradient information on the deep layers when using more loss terms. Therefore, the proposed model obtains better segmentation results.}

In contrast, lines of SPNet to S-7 in Table \ref{tab:Ablation Study on DRIVE dataset.} form a gradually decreasing of local loss functions. When only local loss terms $\mathcal{L}_{l_{2}}$, $\mathcal{L}_{l_{1}}$, and $\mathcal{L}_{l_{0}}$ are used, it obtains high sensitivity but low specificity, which illustrates that the model excessively pays attention to the regions of capillaries and edge vessels. 
With the number of loss functions decreasing, lesser errors are compensated, which causes lower sensitivity.

{In addition, we show the comparison of segmentation results between our SPNet and the U-Net with the proper use of deep supervision. It can be observed that U-Nets with deep supervision (lines of SPNet, S-8, S-9, and S-10) obtain better performance than U-Net (line of backbone) in all metrics. Fortunately, U-Nets with pyramid-like loss (lines of SPNet and S-10) outperform the others (lines of backbone, S-8, and S-9) by a large margin in the metrics of sensitivity and AUC, which verifies the superiority of pyramid-like loss. Moreover, compared with model S-9, model S-8 obtains lower sensitivity; similar results can be observed by comparing SPNet and S-10. This testifies that a $1 \times 1$ convolution operation to produce the side-out prediction maps cannot obtain enough semantic information on the vessel structure. As for the computations of pyramid-like loss, we show the results based on two different strategies, namely down-sampling the ground truth and up-sampling the feature maps to the resolution of the ground truth. It can be observed that there is little difference between the results under the two strategies. To thoroughly analyze the effect of the decoder-sharing mechanism and each term of the loss, we also show the feature maps and segmentation results in Supplementary materials.}

Based on the above observations, the decoder-sharing mechanism and the pyramid-like function are both important, and they help to make great improvement for retinal vessel segmentation.

\subsection{Comparison with state-of-the-art methods}
\label{subsec:Comparison with State-of-the-art Methods}

We present extensive comparisons between SPNet and the state-of-the-art methods, including the works of Azzopardi {\em et al.} \cite{azzopardi2015trainable}, Xie and Tu \cite{xie2015holistically}, DeepVessel \cite{fu2016deepvessel}, Zhou {\em et al.} \cite{zhou2017improving}, Zhao {\em et al.} \cite{zhao2018automatic}, LadderNet \cite{zhuang2018laddernet}, Hu {\em et al.} \cite{hu2018retinal}, Wang {\em et al.} \cite{wang2019blood}, {CS-Net \cite{mou2019cs}, AG-Net \cite{zhang2019attention},} Yu {\em et al.} \cite{yu2020framework}, cross-connected convolutional network (CcNet) \cite{feng2020ccnet}, {NFN$+$ \cite{wu2020nfn+},} and Yang {\em et al.} \cite{yang2021hybrid}. Since our SPNet employs the U-Net \cite{ronneberger2015u} as a backbone, we further make focused comparisons against some variants of U-Net, i.e., multi-scale network followed network model (MS-NFN) \cite{wu2018multiscale}, Yan {\em et al.} \cite{yan2018joint}, and dual encoding U-Net (DEU-Net) \cite{wang2019dual}. The results of LadderNet \cite{zhuang2018laddernet}, Hu {\em et al.} \cite{hu2018retinal}, Wang {\em et al.} \cite{wang2019blood}, Yu {\em et al.} \cite{yu2020framework}, CcNet \cite{feng2020ccnet}, Yang {\em et al.} \cite{yang2021hybrid}, MS-NFN \cite{wu2018multiscale}, Yan {\em et al.} \cite{yan2018joint}, and DEU-Net \cite{wang2019dual} come from the original papers, whereas the others are provided in \cite{mou2020dense}. Since the results on STARE dataset of MS-NFN and DEU-Net are not provided, we implemented the model as described in \cite{wu2018multiscale, wang2019dual}.

\begin{table*}
\caption{{Results on three benchmark datasets. The higher the metric values, the better the method. Our SPNet outperforms the state-of-the-art methods and the variants of U-Net in most performance metrics. The values with bold symbol and underline present the best results and the second best one respectively.}}
\label{tab:Results on three benchmark datasets}
\renewcommand\tabcolsep{2.0pt}
\scriptsize
\begin{center}
\begin{tabular}{l | l|c| c|c|c|c| c|c|c|c| c|c|c|c}
\toprule
\multicolumn{2}{c|}{ } &  & \multicolumn{4}{c|}{DRIVE} & \multicolumn{4}{c|}{STARE} & \multicolumn{4}{c}{CHASE\_DB1}\\
\cline{4-15}
\multicolumn{2}{c|}{Methods} & Year & Acc & Sen & Spe & AUC & Acc & Sen & Spe & AUC & Acc & Sen & Spe & AUC \\
\midrule
\multirow{15}{0.5cm}[0.83cm]{\rotatebox{90}{\parbox{4.2cm}{State-of-the-art Method}}}
& 2nd Human Observer & - & 0.9472 & 0.7760 & 0.9724 &  - & 0.9349 & 0.8952 & 0.9384 &  - & 0.9545 & 0.8105 & 0.9711 &  - \\
& Azzopardi {\em et al.} \cite{azzopardi2015trainable} & 2015 & 0.9442 & 0.7655 & 0.9704 & 0.9614 & 0.9497 & 0.7716 & 0.9701 & 0.9563 & 0.9387 & 0.7585 & 0.9587 &  0.9487 \\
& Xie and Tu \cite{xie2015holistically} & 2015 & 0.9435 & 0.7364 & 0.9730 & 0.9774 & 0.9402 & 0.7116 & 0.9724 & 0.9801 & 0.9380 & 0.7151 & 0.9679 &  0.9798 \\
& DeepVessel \cite{fu2016deepvessel} & 2016 & 0.9533 & 0.7603 & 0.9776 & 0.9789 & 0.9609 & 0.7412 & 0.9701 & 0.9790 & 0.9581 & 0.7130 & 0.9812 &  0.9806 \\
& Zhou {\em et al.} \cite{zhou2017improving} & 2017 & 0.9469 & 0.8078 & 0.9674 & - & 0.9585 & 0.8065 & 0.9761 & - & 0.9520 & 0.7553 & 0.9751 &  - \\
& Zhao {\em et al.} \cite{zhao2018automatic} & 2018 & 0.9580 & 0.7740 & 0.9790 & 0.9750 & 0.9570 & 0.7880 & 0.9760 & 0.9590 &  - &  - &  - &  - \\
& LadderNet \cite{zhuang2018laddernet} & 2018 & 0.9561 & 0.7856 & 0.9810 & 0.9793 & - & - & - & - & 0.9656 & 0.7978 & 0.9818 &  0.9839 \\
& Hu {\em et al.} \cite{hu2018retinal} & 2018 & 0.9533 & 0.7772 & 0.9793 & 0.9759 & 0.9632 & 0.7543 & 0.9814 & 0.9751 & - & - & - & - \\
& Wang {\em et al.} \cite{wang2019blood} & 2019 & 0.9541 & 0.7648 & 0.9817 & - & 0.9640 & 0.7523 & \textbf{0.9885} & - & 0.9603 & 0.7730 & 0.9792 & - \\
& CS-Net \cite{mou2019cs} & 2019 & 0.9632 & 0.8170 & \textbf{0.9854} & 0.9798 & \textbf{0.9752} & \textbf{0.8816} & 0.9840 & \textbf{0.9932} & - & - & - & - \\
& AG-Net \cite{zhang2019attention} & 2019 & \textbf{0.9692} & 0.8100 &  \underline{0.9848} & \textbf{0.9856} & - & - & - & - & \textbf{0.9743} &  \underline{0.8186} &  \underline{0.9848} &  \underline{0.9863} \\
& Yu {\em et al.} \cite{yu2020framework} & 2020 & 0.9524 & 0.7643 & 0.9803 & 0.9723 & 0.9613 & 0.7837 & 0.9822 & 0.9787 & - & - & - & - \\
& CcNet \cite{feng2020ccnet} & 2020 & 0.9528 & 0.7625 & 0.9809 & 0.9678 & 0.9633 & 0.7709 & 0.9848 & 0.9700 & - & - & - & - \\
& NFN$+$ \cite{wu2020nfn+} & 2020 & 0.9582 & 0.7996 & 0.9813 &  \underline{0.9830} & 0.9672 & 0.7963 &  \underline{0.9863} &  \underline{0.9875} & \underline{0.9688} & 0.8003 & \textbf{0.9880} & \textbf{0.9894} \\
& Yang {\em et al.} \cite{yang2021hybrid} & 2021 & 0.9579 & \textbf{0.8353} & 0.9751 & - & 0.9626 & 0.7946 & 0.9821 & - & 0.9632 & 0.8176 & 0.9776 & -  \\
& \textbf{SPNet (ours)} & - & \underline{0.9664} & \underline{0.8243} & 0.9802 & 0.9828 &  \underline{0.9692} & \underline{0.8504} & 0.9790 & 0.9812 & 0.9685 &  \textbf{0.8619} & 0.9760 &  0.9840 \\
\hline
\hline
\multirow{15}{0.5cm}[2.53cm]{\rotatebox{90}{\parbox{3.0cm}{Variant of U-Net}}}
& Backbone (U-Net \cite{ronneberger2015u}) & 2015 & 0.9501 & 0.7356 & 0.9602 & 0.9641 & 0.9517 & 0.7101 & 0.9682 & 0.9615 & 0.9499 & 0.7094 & 0.9767 &  0.9613 \\
& MS-NFN \cite{wu2018multiscale} & 2018 & 0.9567 & 0.7844 & \textbf{0.9819} & 0.9807 & 0.9655 & 0.7688 & 0.9823 & 0.9678 & 0.9637 & 0.7538 & \textbf{0.9847} & 0.9825 \\
& Yan {\em et al.} \cite{yan2018joint} & 2018 & 0.9542 & 0.7653 & 0.9818 & 0.9752 & 0.9612 & 0.7581 & \textbf{0.9846} & 0.9801 & 0.9610 & 0.7633 & 0.9809 & 0.9781 \\
& DEU-Net \cite{wang2019dual} & 2019 & 0.9567 & 0.7940 & 0.9816 & 0.9772 & 0.9652 & 0.7614 & 0.9818 & 0.9759 & 0.9661 & 0.8074 &  0.9821 &  0.9812 \\
& \textbf{SPNet (ours)} & - &  \textbf{0.9664} & \textbf{0.8243} & 0.9802 & \textbf{0.9828} & \textbf{0.9692} &  \textbf{0.8504} & 0.9790 & \textbf{0.9812} &  \textbf{0.9685} &  \textbf{0.8619} & 0.9760 &  \textbf{0.9840} \\
\bottomrule
\end{tabular}
\end{center}
\end{table*}

\begin{figure*}[t]
\begin{center}
\includegraphics[width=1.0\linewidth]{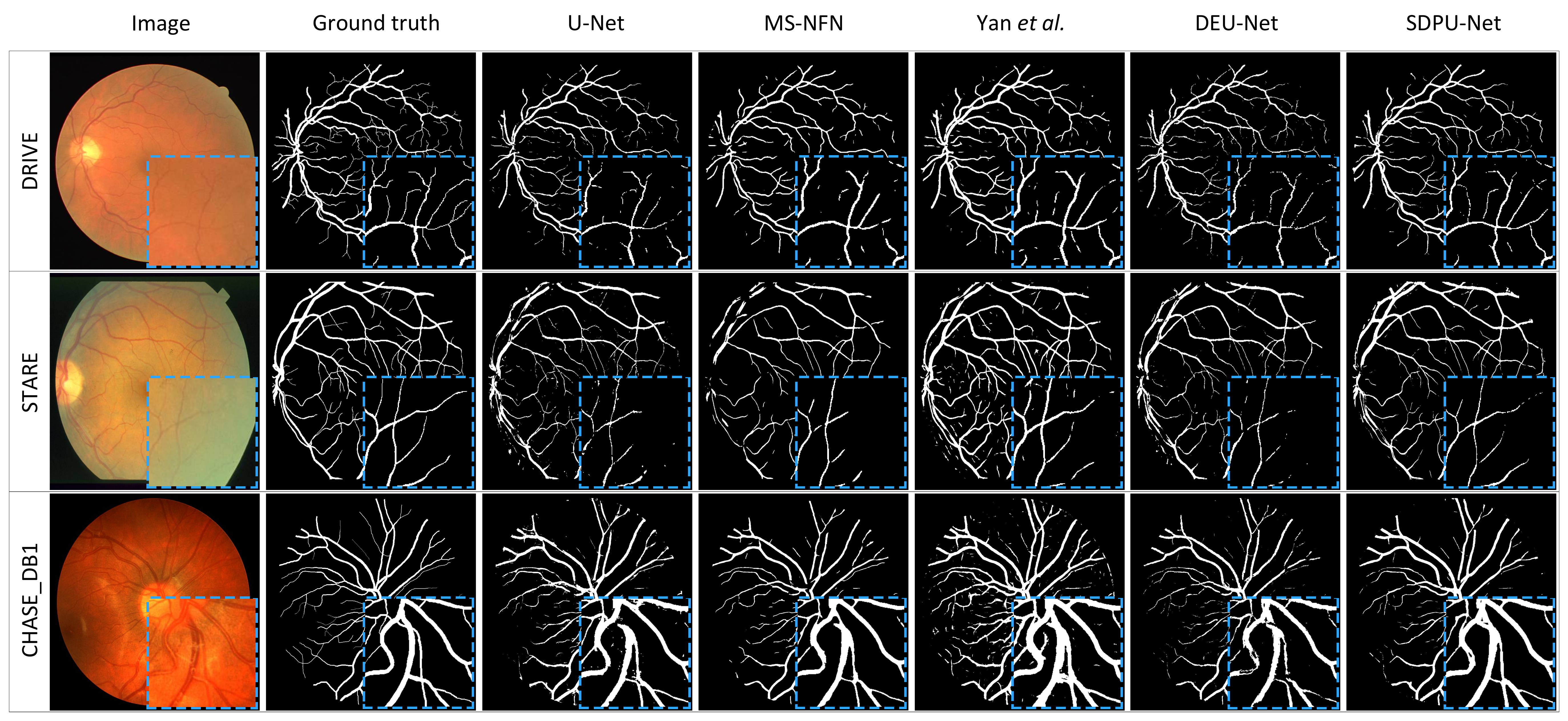}
\end{center}
   \caption{The segmentation results on DRIVE, STARE, and CHASE\_DB1 datasets. From left to right: test image, ground truth, the segmentation results obtained by U-Net \cite{ronneberger2015u}, MS-NFN \cite{wu2018multiscale}, Yan {\em et al.} \cite{yan2018joint}, DEU-Net \cite{wang2019dual}, and our SPNet. A zoom-in visualization is also provided with a dashed box. Due to variations and blurry boundaries of vessels, U-Net, MS-NFN, and DEU-Net fail to extract the contours and some thin and tiny vessels. However, our method achieves superior segmentation results. (Best viewed in color.)}
\label{fig:The segmentation results on DRIVE and CHASE_DB1 datasets.}
\end{figure*}

\begin{figure*}
    \centering
    \subfigure[ROC on DRIVE]{
        \label{ROC on DRIVE}
        \begin{minipage}[b]{0.32\textwidth}
        \includegraphics[width=1\textwidth]{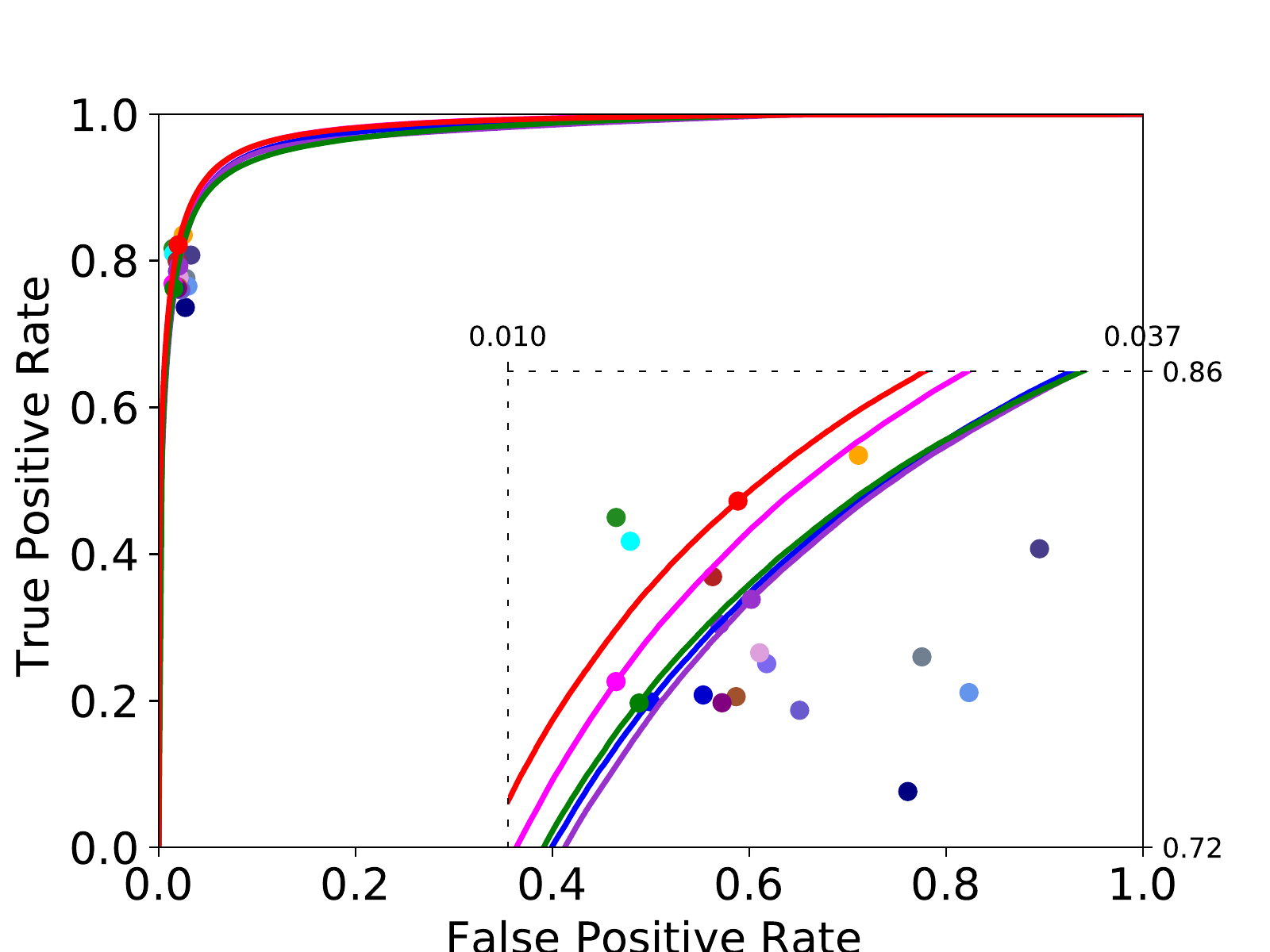}
        \end{minipage}}
    \subfigure[ROC on STARE]{
        \label{ROC on STARE}
        \begin{minipage}[b]{0.32\textwidth}
        \includegraphics[width=1\textwidth]{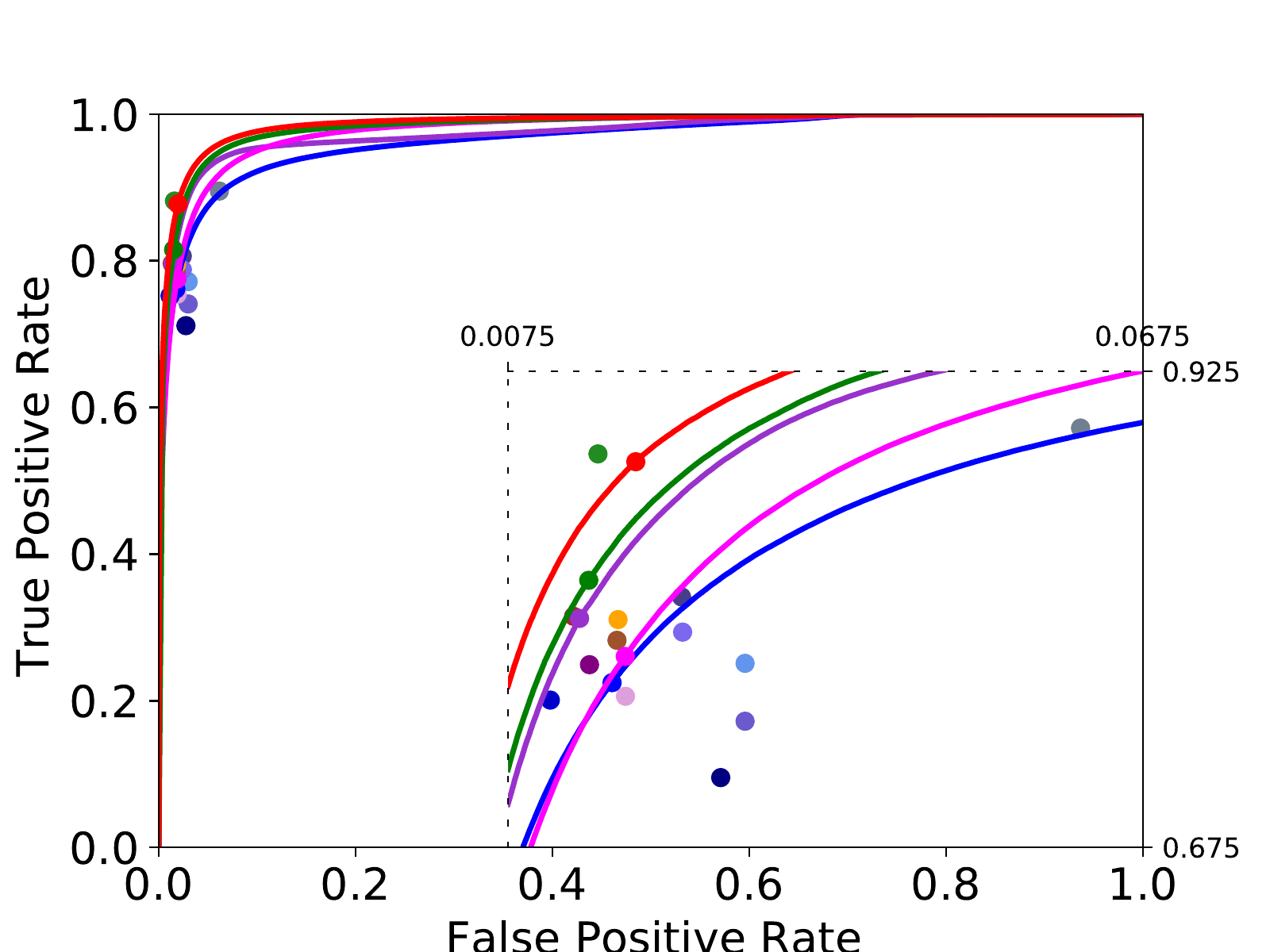}
        \end{minipage}}
    \subfigure[ROC on CHASE\_DB1]{
        \label{ROC on CHASE_DB1}
        \begin{minipage}[b]{0.32\textwidth}
        \includegraphics[width=1\textwidth]{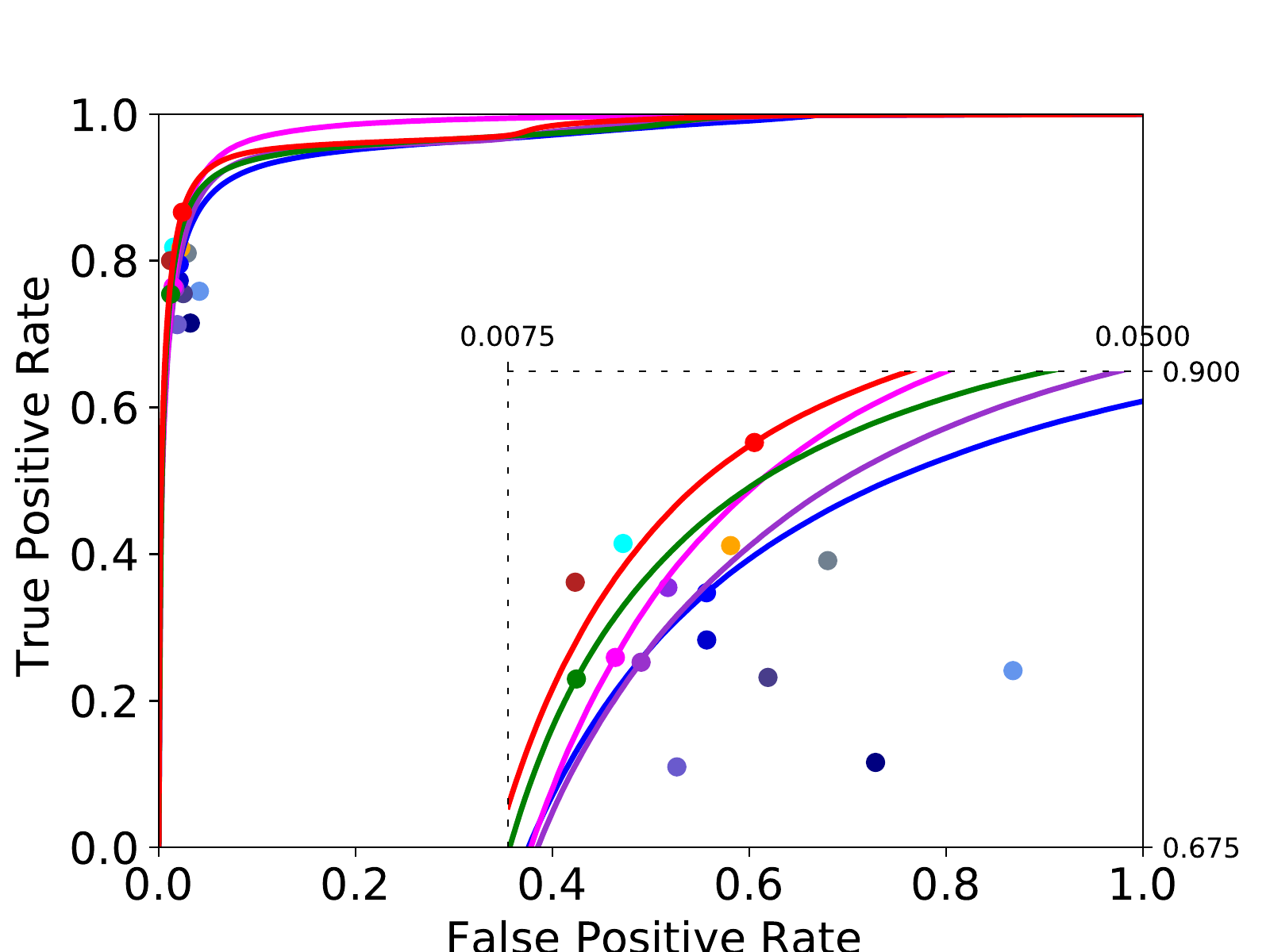}
        \end{minipage}}
    \subfigure[P-R Curve on DRIVE]{
        \label{P-R Curve on DRIVE}
        \begin{minipage}[b]{0.32\textwidth}
        \includegraphics[width=1\textwidth]{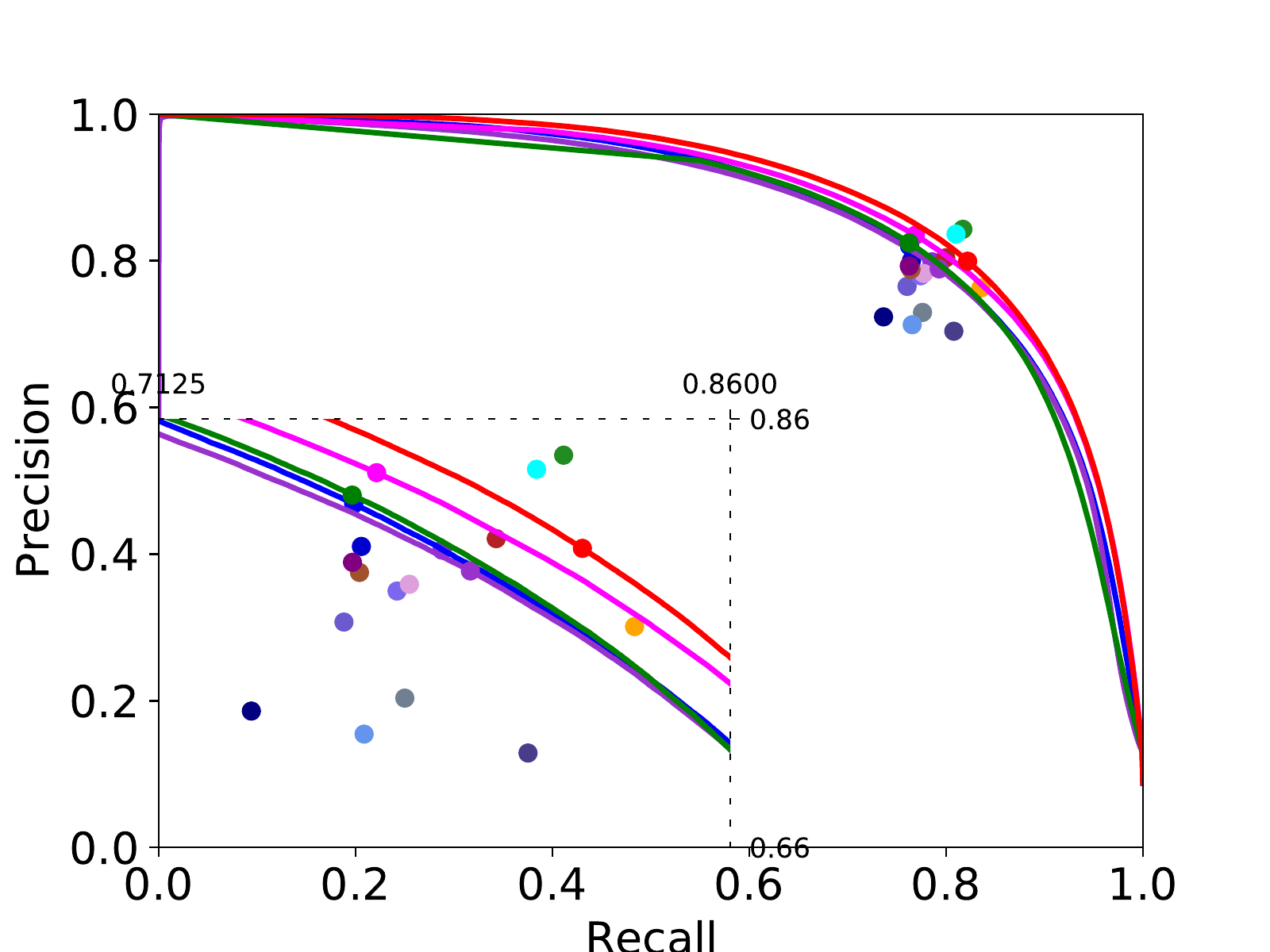}
        \end{minipage}}
    \subfigure[P-R Curve on STARE]{
        \label{P-R Curve on STARE}
        \begin{minipage}[b]{0.32\textwidth}
        \includegraphics[width=1\textwidth]{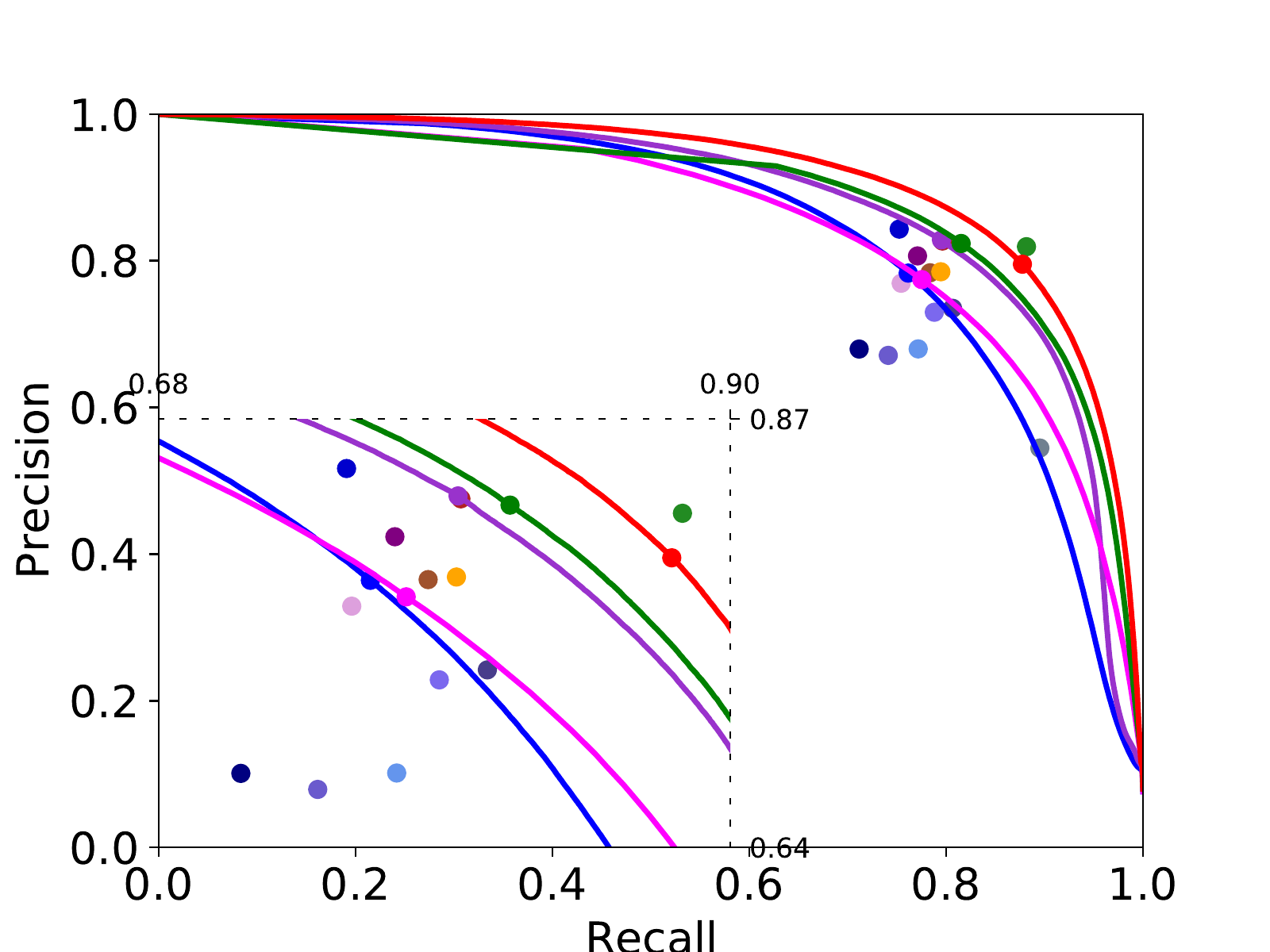}
        \end{minipage}}
    \subfigure[P-R Curve on CHASE\_DB1]{
        \label{P-R Curve on CHASE_DB1}
        \begin{minipage}[b]{0.32\textwidth}
        \includegraphics[width=1\textwidth]{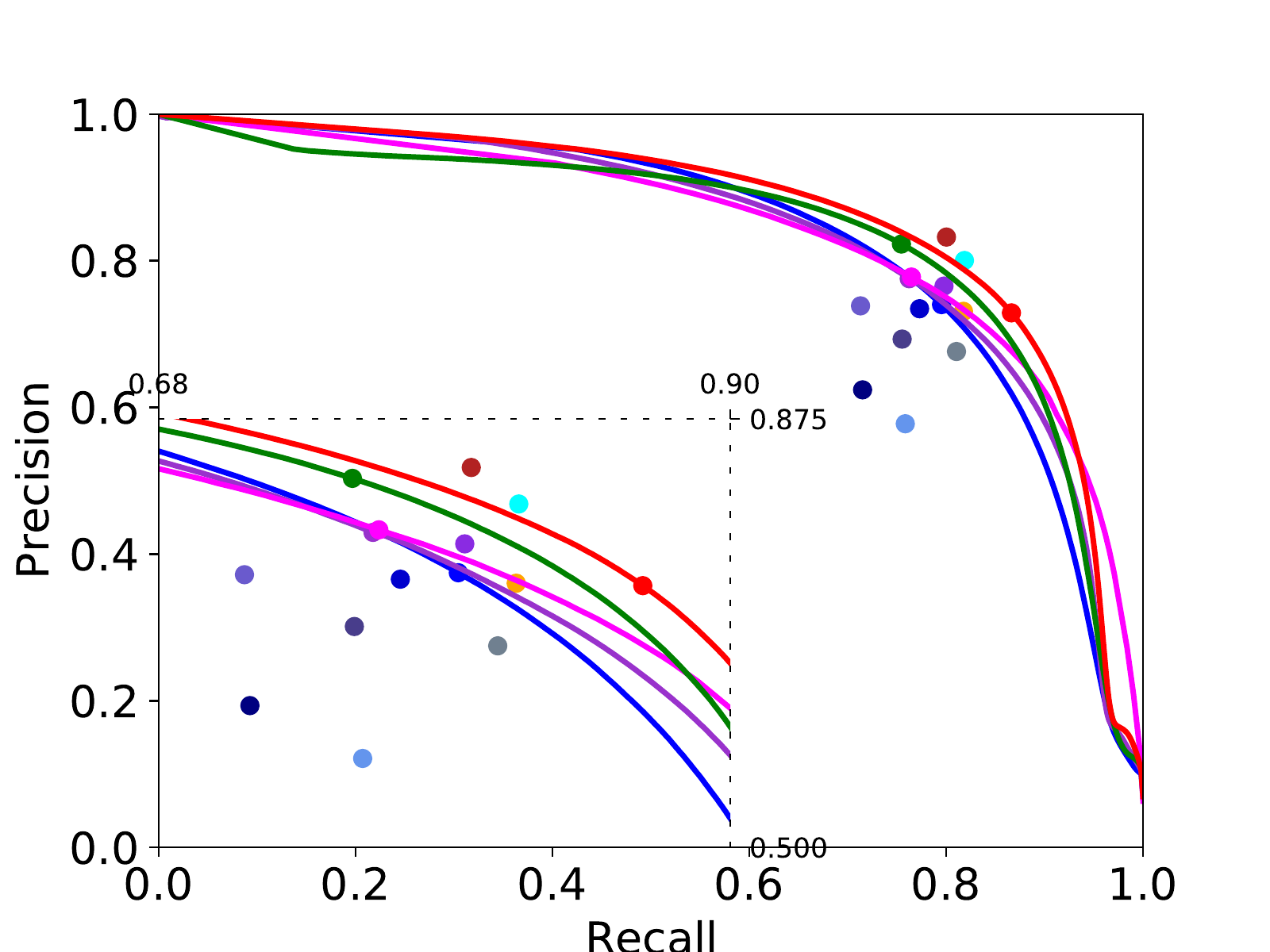}
        \end{minipage}}
    \subfigure{
        \label{f1}
        \begin{minipage}[b]{0.99\textwidth}
        \includegraphics[width=1\textwidth,trim=126 30 92 84,clip]{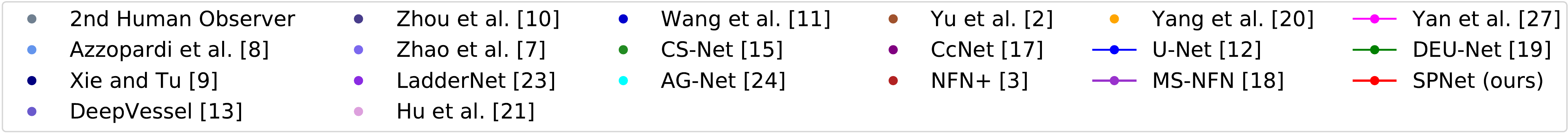}
        \end{minipage}}
    \caption{Receiver operating characteristic (ROC) and Precision-Recall (P-R) curves on the three datasets. The points indicate the hard segmentation results. A zoom-in visualization is provided with a dashed box. It can be observed that our SPNet outperforms other methods not only on the hard segmentation point, but also the whole curve. (Best viewed in color.)}
    \label{fig:plot_ROC_PR.}
\end{figure*}

\begin{figure*}
\begin{center}
\includegraphics[width=0.98\linewidth]{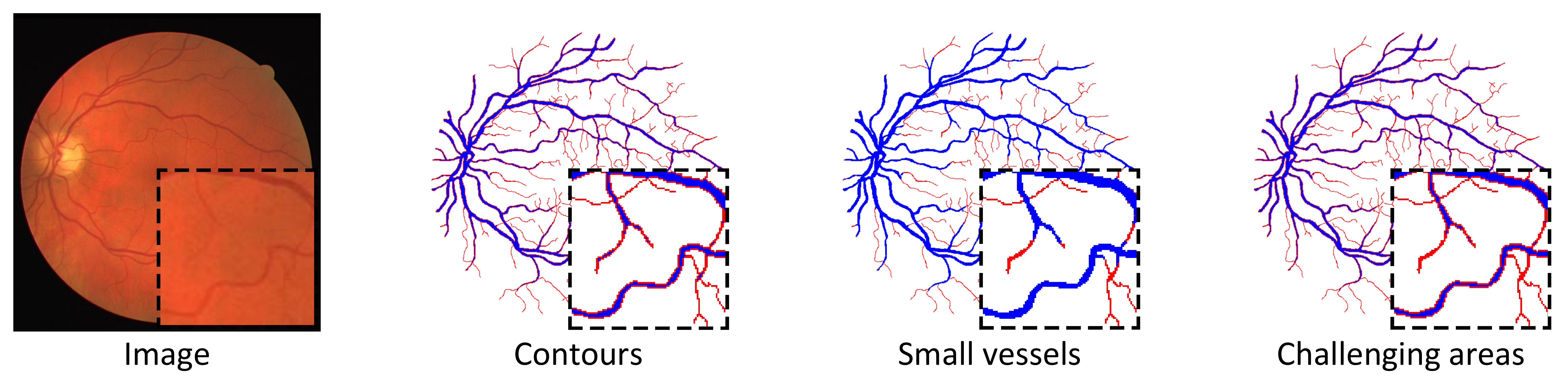}
\end{center}
   \caption{Example of morphological operations to generate the contours and small vessels. For each retinal image (1st column), we mark the contours (2nd column) and the small vessels (3rd column) in red separately, and then execute a union operation to obtain the challenging areas (4th column). (Best viewed in color.)}
\label{fig:morphological operations.}
\end{figure*}

Quantitative results with respect to the state-of-the-art methods are shown at the top lines of Table \ref{tab:Results on three benchmark datasets}. Our SPNet outperforms the other methods in most performance metrics. On DRIVE dataset, CS-Net \cite{mou2019cs} and Yang {\em et al.} \cite{yang2021hybrid} obtain better performance in terms of specificity and sensitivity, respectively, but SPNet outperforms them when evaluated by the other metrics by a large margin. Compared with AG-Net \cite{zhang2019attention}, which uses the multi-scale loss as the auxiliary loss for U-Net, SPNet obtains better performance in terms of sensitivity. This verifies the superiority of pyramid-like loss. 
Note that there is not a method performing best in all metrics. Fortunately, SPNet yields competitive results in accuracy, sensitivity, and AUC with those single-metric best results, which shows that our method accurately classifies most pixels including blood vessels and background. We attribute this effectiveness to the use of SDMs, by which the model shares the knowledge learned from multi-scale features and consequently promotes the overall segmentation result.

On STARE dataset, although Wang {\em et al.} \cite{wang2019blood} performs best in terms of specificity, but it obtains a sensitivity of 0.7523, which is relatively poor among all results. Compared with the other methods, our SPNet and CS-Net \cite{mou2019cs} perform much better by sensitivity measure, while maintaining better or competitive results in accuracy, specificity, and AUC. In particular, SPNet obtains a sensitivity of 0.8504, which approaches the second expert’s manual segmentation. This verifies the effectiveness of our SPNet.

Similarly, on CHASE\_DB1 dataset, our SPNet obtains competitive results, especially in terms of sensitivity. In contrast to the first two benchmark datasets, retinal images in CHASE\_DB1 have higher resolution and suffer from unbalanced illumination and thicker arterioles. However, our SPNet still obtains a sensitivity of 0.8619, which outperforms other methods by 0.0433 at least. This demonstrates that the integration of SDMs and pyramid-like loss helps the model to strengthen characterization on the challenging areas, namely the small and low intensity contrast vessels. Therefore, more blood vessels in the retinal images are successfully extracted by our SPNet.

At the bottom lines of Table \ref{tab:Results on three benchmark datasets}, we specifically compare our SPNet with the U-Net and some variants. We summarize some conclusions as follows. First, our SPNet outperforms the backbone by a large margin in all metrics, which verifies the effectiveness of the proposed decoder-sharing mechanism and pyramid-like loss function. Second, compared with MS-NFN \cite{wu2018multiscale} and Yan {\em et al.} \cite{yan2018joint}, which are also specially designed for addressing the problem of thickness inconsistency, SPNet obtains higher sensitivity on all three datasets. This demonstrates that the majority of blood vessels are extracted successfully by our SPNet. Finally, high accuracy and AUC show that SPNet has an overall segmentation performance.

Some test results are presented in Fig. \ref{fig:The segmentation results on DRIVE and CHASE_DB1 datasets.}. It can be observed that U-Net \cite{ronneberger2015u}, the multi-scale method MS-NFN \cite{wu2018multiscale}, the joint-loss framework of Yan {\em et al.} \cite{yan2018joint}, and the DEU-Net \cite{wang2019dual} fail to extract most thin and edge vessels. However, our SPNet obtains distinct segmentation results despite the variations and blurry boundaries of vessels in fundus images.

In addition, the receiver operating characteristic (ROC) and Precision-Recall (P-R) curves are shown in Fig. \ref{fig:plot_ROC_PR.}. We plot the curves for U-Net \cite{ronneberger2015u}, MS-NFN \cite{wu2018multiscale}, Yan {\em et al.} \cite{yan2018joint}, DEU-Net \cite{wang2019dual}, and our SPNet, and the hard segmentation results for the other methods. We also display a zoom-in visualization in this figure. It can be observed that our method outperforms the others not only on the hard segmentation point, but also the whole curve.

\subsection{Evaluation on contours and small vessels}
\label{subsec:Evaluation on Contours and Small Vessels}

So far we have evaluated the segmentation results in an overall level. To further analyze the segmentation performance, we use morphological operations to extract the challenging areas: the contours and the small vessels (CS vessels for short). Fig. \ref{fig:morphological operations.} plots some examples generated by the referred morphological operations. First, we use the erosion operation based on a disk-shaped structure element with a radius of 1 pixel to obtain the centerline of blood vessels. Those removed parts are referred to as contours. Second, we implement the opening operation and the closing operation based on a disk-shaped structure element with a radius of 1 and 8 pixels respectively. We also remove the objects smaller than 100 pixels based on a 4-neighborhood connectivity. The remaining vessels are referred to as big vessels and the others are small ones. Lastly, we use a union operation on the two parts (the contours and the small vessels) to obtain the challenging areas.

\begin{table}[!t]
\caption{Segmentation performance evaluation on the contours and small vessels (CS vessels for short) versus the non-CS vessels.}
\label{tab:Segmentation performance evaluation on the contours and small vessels}
\small
\renewcommand\tabcolsep{3.0pt}
\begin{center}
\begin{tabular}{l|c|c|c|c|c|c}
\toprule
\multicolumn{2}{c|}{} & U-Net & MS-NFN & Yan {\em et al.} & DEU-Net & \textbf{SPNet} \\
\multicolumn{2}{c|}{} & \cite{ronneberger2015u} & \cite{wu2018multiscale} & \cite{yan2018joint} & \cite{wang2019dual} & \textbf{(ours)} \\
\midrule
\multicolumn{7}{c}{CS vessels} \\
DRIVE & Acc & 0.9759 & 0.9734 & \textbf{0.9783} & 0.9763 & 0.9748 \\
      & Sen & 0.6710 & 0.7199 & 0.7114 & 0.6723 & \textbf{0.7526} \\
      & Spe & 0.9849 & 0.9809 & \textbf{0.9863} & 0.9854 & 0.9814 \\
      & AUC & 0.9803 & 0.9791 & 0.9770 & 0.9723 & \textbf{0.9804} \\
STARE & Acc & 0.9706 & 0.9747 & \textbf{0.9749} & 0.9738 & 0.9744 \\
      & Sen & 0.6932 & 0.6413 & 0.6745 & 0.6292 & \textbf{0.7534} \\
      & Spe & 0.9777 & \textbf{0.9833} & 0.9824 & 0.9828 & 0.9801 \\
      & AUC & 0.9674 & 0.9649 & 0.9757 & 0.9683 & \textbf{0.9738} \\  
CHASE\_DB1 & Acc & 0.9746 & 0.9777 & 0.9797 & \textbf{0.9812} & 0.9737 \\
           & Sen & 0.6305 & 0.5736 & 0.6072 & 0.5409 & \textbf{0.7535} \\
           & Spe & 0.9803 & 0.9845 & 0.9860 & \textbf{0.9886} & 0.9773 \\
           & AUC & 0.9563 & 0.9621 & 0.9701 & 0.9693 & \textbf{0.9745} \\
\multicolumn{7}{c}{non-CS vessels} \\
DRIVE & Acc & 0.9825 & 0.9788 & \textbf{0.9841} & 0.9831 & 0.9803 \\
      & Sen & 0.9338 & 0.9387 & 0.9401 & 0.9391 & \textbf{0.9616} \\
      & Spe & 0.9847 & 0.9806 & 0.9825 & \textbf{0.9852} & 0.9812 \\
      & AUC & 0.9957 & 0.9942 & 0.9893 & 0.9941 & \textbf{0.9967} \\
STARE & Acc & 0.9733 & \textbf{0.9781} & 0.9769 & 0.9776 & 0.9778 \\
      & Sen & 0.8891 & 0.8760 & 0.8624 & 0.8756 & \textbf{0.9358} \\
      & Spe & 0.9773 & \textbf{0.9829} & 0.9821 & 0.9824 & 0.9798 \\
      & AUC & 0.9849 & 0.9814 & 0.9810 & 0.9868 & \textbf{0.9914} \\    
CHASE\_DB1 & Acc & 0.9748 & 0.9776 & 0.9801 & \textbf{0.9817} & 0.9744 \\
           & Sen & 0.8824 & 0.8566 & 0.8605 & 0.8584 & \textbf{0.9290} \\
           & Spe & 0.9797 & 0.9840 & 0.9856 & \textbf{0.9882} & 0.9765 \\
           & AUC & 0.9830 & 0.9841 & 0.9859 & 0.9888 & \textbf{0.9911} \\
\bottomrule
\end{tabular}
\end{center}
\end{table}

Table \ref{tab:Segmentation performance evaluation on the contours and small vessels} shows the results on the CS vessels versus the non-CS ones. It shows that U-Net \cite{ronneberger2015u}, MS-NFN \cite{wu2018multiscale}, Yan {\em et al.} \cite{yan2018joint}, and DEU-Net \cite{wang2019dual} tend to give more attention to the background and obtain higher specificity. These methods also get slightly higher accuracy since the background occupies the vast majority of pixels in the whole retinal image. For the non-CS vessels, all methods give acceptable results. For the CS vessels, the SPNet achieves obviously higher sensitivity and AUC. This verifies that our method can effectively learn the semantic information and obtain accurate segmentation for the challenging areas.

\begin{table}[!t]
\caption{Evaluation of vessel structure using the CAL.}
\label{tab:Evaluation of vessel structure using the CAL}
\begin{center}
\begin{tabular}{l|c|c|c|c}
\toprule
Method & $C$ & $A$ & $L$ & $CAL$ \\
\midrule
\multicolumn{5}{c}{DRIVE} \\
U-Net \cite{ronneberger2015u} & 0.9934 & 0.9262 & 0.8734 & 0.8042 \\
MS-NFN \cite{wu2018multiscale} & \textbf{0.9975} & 0.9168 & 0.8595 & 0.7867 \\
Yan {\em et al.} \cite{yan2018joint} & 0.9967 & 0.9070 & 0.8332 & 0.7544 \\
DEU-Net \cite{wang2019dual}   & 0.9889 & \textbf{0.9362} & \textbf{0.8867} & 0.8212 \\
\textbf{SPNet (ours)}        & 0.9935 & 0.9353 & 0.8852 & \textbf{0.8228} \\
\multicolumn{5}{c}{STARE} \\
U-Net \cite{ronneberger2015u} & 0.9892 & 0.8611 & 0.8304 & 0.7097 \\
MS-NFN \cite{wu2018multiscale} & \textbf{0.9981} & 0.8434 & 0.8333 & 0.7080 \\
Yan {\em et al.} \cite{yan2018joint} & 0.9949 & 0.8542 & 0.8366 & 0.7178 \\
DEU-Net \cite{wang2019dual}   & 0.9913 & 0.8571 & 0.8341 & 0.7129 \\
\textbf{SPNet (ours)}        & 0.9926 & \textbf{0.8954} & \textbf{0.8633} & \textbf{0.7702} \\
\multicolumn{5}{c}{CHASE\_DB1} \\
U-Net \cite{ronneberger2015u} & 0.9950 & 0.8467 & 0.8112 & 0.6843 \\
MS-NFN \cite{wu2018multiscale} & \textbf{0.9990} & 0.8336 & 0.8152 & 0.6806 \\
Yan {\em et al.} \cite{yan2018joint} & 0.9978 & 0.8857 & 0.8287 & 0.7086 \\
DEU-Net \cite{wang2019dual}   & 0.9967 & 0.8765 & 0.8500 & 0.7432 \\
\textbf{SPNet (ours)}        & 0.9977 & \textbf{0.8794} & \textbf{0.8553} & \textbf{0.7511} \\
\bottomrule
\end{tabular}
\end{center}
\end{table}

\subsection{Evaluation of vessel structure using the CAL}
\label{subsec:Evaluation of Vessel Structure Using the CAL}

In addition to the metrics sensitivity and specificity, which are based on pixel-wise matching strategy, we implement $CAL$ \cite{gegundez2012function} to assess the vascular structure. To be specific, the quality evaluation function for vessel segmentation contains three factors, i.e., connectivity ($C$), area ($A$), and length ($L$). It can be formally expressed as
\begin{equation}
f(C, A, L) = C \cdot A \cdot L \equiv CAL.
\end{equation}

Table \ref{tab:Evaluation of vessel structure using the CAL} shows the segmentation performance measured by $CAL$. The comparable performance on parameter $C$ reflects that our SPNet obtains less fragmented segmentations. The comparable even better performance on parameter $A$ presents that the generated segmentation maps of our method have more overlapping areas with the reference manual annotations. Furthermore, the comparable even better performance on parameter $L$ shows that our SPNet has outstanding capability in detecting the lengths of the skeletons. Overall, our method achieves 0.8228, 0.7242, and 0.7511 in terms of $CAL$, outperforming the others in the three fundus datasets.

\subsection{Evaluation on cross-datasets}
\label{subsec:Evaluation on Cross-datasets}

In practice, visual discrepancies among the clinical datasets could be large \cite{ren2021learning, ren2022multi, xu2022cross, ren2021heterogeneous}. For example, DRIVE dataset contains 33 images without any sign of diabetic retinopathy and 7 images with signs of mild early diabetic retinopathy. STARE dataset is visually similar to DRIVE, but with more discoloration of the optic nerve. Compared with these two datasets, the photographs in CHASE\_DB1 dataset have higher resolution, but with unbalanced illumination. It requires the segmentation method to be robust on various retinal fundus images. Here we conduct a group of experiments under the setting of cross-dataset to demonstrate the robustness of our SPNet. In other words, we train the models on one dataset and test on another.

The performance of AUC is shown in Table \ref{tab:AUC performance on cross-dataset}. We see that our SPNet obtains the best results on 5 out of 6 tasks. On the tasks of D$\rightarrow$S and S$\rightarrow$D, our SPNet achieves 0.9734 and 0.9599 respectively, which are close to or even exceed the results of some supervised methods in Table \ref{tab:Results on three benchmark datasets}. Although the tasks of D$\rightarrow$C and S$\rightarrow$C are relatively hard, SPNet still obtains higher AUC compared with the other methods. In the last two tasks (C$\rightarrow$D and C$\rightarrow$S), SPNet outperforms others by a large margin. These improvements can be attributed to the help of SDM, by which the blood vessels at various scales can be successfully extracted from fundus images. Hence, our SPNet shows stronger generalization ability than the other methods.

\begin{table}
\caption{AUC performance on cross-dataset. The tasks are denoted as ``training dataset $\rightarrow$ testing dataset". (D: DRIVE; S: STARE; C: CHASE\_DB1.)}
\label{tab:AUC performance on cross-dataset}
\begin{center}
\begin{tabular}{l|c|c|c|c|c}
\toprule
Task & U-Net & MS-NFN & Yan {\em et al.} & DEU-Net & \textbf{SPNet} \\
  & \cite{ronneberger2015u} & \cite{wu2018multiscale} & \cite{yan2018joint} & \cite{wang2019dual} & \textbf{(ours)} \\
\midrule
D $\rightarrow$ S & 0.9384 & 0.9558 & 0.9708 & 0.9511 & \textbf{0.9734} \\
D $\rightarrow$ C & 0.7920 & 0.8717 &  -         & 0.8687 & \textbf{0.8995} \\
S $\rightarrow$ D & 0.9157 & 0.9459 & \textbf{0.9599} & 0.9569 & \textbf{0.9599} \\
S $\rightarrow$ C & 0.8420 & 0.8952 &  -         & \textbf{0.9068} & 0.9043 \\
C $\rightarrow$ D & 0.9265 & 0.9399 &  -         & 0.9471 & \textbf{0.9650} \\
C $\rightarrow$ S & 0.9406 & 0.9436 &  -         & 0.9576 & \textbf{0.9744} \\
\bottomrule
\end{tabular}
\end{center}
\end{table}

\subsection{Training and test complexity}
\label{subsec:Training and Test Complexity}

Owing to the design of SDM, the proposed network requires a lower computational cost than the other variants of U-Net. Specifically, in the decoder-sharing mechanism, multi-scale features are decoded simultaneously without introducing additional parameters. However, MS-NFN \cite{wu2018multiscale} and DEU-Net \cite{wang2019dual} apply many other modules to generate multi-scale feature maps. On DRIVE dataset, MS-NFN and DEU-Net take more than 16 hours and more than 13 hours respectively to train the models, whereas our SPNet takes 1.80 hours to train the model (Intel Xeon E5-2699 v3 CPU @2.30GHz, NVIDIA Titan Xp GPU, 128 GB Memory, and PyTorch 1.2.0). On CHASE\_DB1 dataset, it takes more than 30 hours to train MS-NFN, and more than 20 hours to train DEU-Net, whereas 3.88 hours to train our SPNet. During applying the trained neural network, SPNet takes only 1.26 seconds to segment a $565 \times 584$ retinal image on DRIVE dataset and 3.83 seconds to segment a $999 \times 960$ retinal image on CHASE\_DB1 dataset, whereas DEU-Net takes 6.75 seconds and 12.82 seconds respectively. Hence, SPNet has lower computational complexity than MS-NFN and DEU-Net. These verify that the proposed method is competitive not just in effectiveness but also in efficiency.

\section{Discussion}
\label{sec:Discussion}

In this section, we further analyze and discuss our work in terms of the dataset, method, and some important experimental results.

Retinal vessel segmentation in color fundus images is a meaningful and challenging task. Among most related works, there are three commonly used benchmark datasets, namely DRIVE, STARE, and CHASE\_DB1. These datasets consist of fewer images compared with the datasets for other segmentation tasks, e.g. COCO\footnote{https://cocodataset.org}. This brings a problem of learning from relatively limited training data. To address this challenge, we randomly crop patches from the original images for data augmentation and train our SPNet on these samples. The segmentation results on these datasets are reported in most state-of-the-art deep learning-based networks, which enables us to make a direct comparison with them. Future evaluations of our SPNet on other real-world datasets including HRF \cite{budai2013robust} and ROSE \cite{ma2021rose} datasets will be our future works. In addition, the data shift (or domain shift) among these datasets cannot be ignored. Therefore, extending our method to the domain adaptation scenario \cite{ren2020domain, luo2022unsupervised, ren2020generalized, luo2021conditional} helps us to improve the generalization ability.

For vessel extraction task, the backbones (i.e., basic neural networks) of deep learning-based methods have been developed from original FCNs to U-shaped architectures, making progress in corresponding segmentation performance. However, one of the disadvantages of U-Net is its inadequate ability in extracting multi-scale features. To alleviate this problem, we propose a decoder-sharing mechanism to learn multi-scale contextual information simultaneously. Experimental results show that this mechanism achieves great improvement compared with the backbone U-Net. Recently, there are several inspiring methods that are proposed to further improve the segmentation performance. For instance, {Wu {\em et al.} \cite{wu2020nfn+} propose a novel network followed network for retinal vessel segmentation, and employ the inter-network skip connections to make the most of deep feature maps. Mou {\em et al.} \cite{mou2019cs} propose a self-attention mechanism to aggregate the global contextual information and thus improve the segmentation performance for retinal vessel. To filter out the noise in the background and preserve the semantic information in the boundary, Zhang {\em et al.} \cite{zhang2019attention} propose an attention guided filter as an expanding path.} 
Zhou {\em et al.} \cite{zhou2021refined} design a symmetric equilibrium generative adversarial network to enhance the performance of a single U-Net. Cherukuri {\em et al.} \cite{cherukuri2020deep} extract image patterns and noisy patches before training phase and use a residual network architecture \cite{he2016deep} as the backbone. An interesting research direction is to combine these methods with our SPNet and further achieve the segmentation improvement on the contours and the small vessels.

The experimental results show that our SPNet obtains better results than the state-of-the-art methods in most performance metrics. Note that the non-vessel background occupies the vast majority of pixels in the whole retinal image. Thus, small increases in the performance metrics may indicate large improvement in the vessel extraction. In order to quantitatively investigate the effectiveness of our method in segmenting the thin and edge vessels, we specially extract these challenging areas and evaluate the corresponding performance in Section \ref{subsec:Evaluation on Contours and Small Vessels}. The results show that our SPNet performs much better than some recent variants of U-Net in the challenging areas, while maintaining the performance in the other areas.

\section{Conclusion}
\label{sec:Conclusion}

In this paper, we present a novel deep neural network, named SPNet, for retinal blood vessel segmentation. Specifically, a decoder-sharing mechanism is introduced to capture the multi-scale semantic information, and the residual pyramid architecture is designed to help the end-to-end network to compensate possible segmentation errors progressively in the decoder. Integration of these two modules makes the learning process of semantic information in our method look like an error-compensation fashion.

Compared with the previous methods, the main advantages of our SPNet are as follows. 
First, without introducing extra network parameters, we integrate the proposed decoder-sharing mechanism and pyramid-like loss with U-shaped architectures. These variants improve segmentation performance on the contours and the small vessels by a large margin. 
Second, we overcome the problem of various thicknesses in an intuitively understandable model, rather than an ensemble of multiple sub-models. 
Third, in terms of effectiveness and efficiency, the proposed SPNet achieves outstanding performance consistently on three public retinal vessel segmentation datasets. In addition, performances on cross-datasets verify that our SPNet shows stronger generalization ability.

The proposed method is currently applied to retinal blood vessel segmentation. However, we believe that it can accurately and efficiently extract blood vessels in other structures and organs, which also suffer from challenges of blurry boundaries and scale variation. How to extend the method to process more clinical angiography data, as well as 3D OCT images, is our future work.

\section*{Acknowledgements}

This work is supported in part by the National Natural Science Foundation of China under Grants 61976229, 61906046.

\bibliography{SPNet_bib}

\end{document}